\documentclass[english]{article}
\usepackage[T1]{fontenc}
\usepackage[latin9]{inputenc}
\usepackage{geometry}
\usepackage{array}
\usepackage{textcomp}
\usepackage{amsmath}
\usepackage{amssymb}
\usepackage{graphicx}
\usepackage{setspace}
\makeatletter

\providecommand{\tabularnewline}{\\}

\usepackage{indentfirst}

\makeatother

\usepackage{babel}
\begin{document}
\title{Mixing Constant Sum and Constant Product Market Makers}
\author{Alexander Port\,\,\,\,\,Neelesh Tiruviluamala\thanks{Emails: alex@thrackle.io, neel@thrackle.io}}
\maketitle
\begin{abstract}
Two popular forms of automated market makers are constant sum and
constant product (CSMM and CPMM respectively). Each has its advantages
and disadvantages: CSMMs have stable exchange rates but are vulnerable
to arbitrage and can sometimes fail to provide liquidity, while a
CPMM can have large impermanent loss due to exchange rate changes
but are always able to provide liquidity to participants.

A significant amount of work has been done in order to get the best
of both constant sum and constant product characteristics. Perhaps
most the relevant to this paper is Stableswap, which has an ``amplification
coefficient'' parameter controlling the balance between the two types
of behavior \cite{Ego19}. Alternative approaches, such as in \cite{AEC21},
involve constructing AMMs using portfolio value functions. However,
there is still much work to be done on these fronts. This paper presents
multiple novel methods for mixing market makers and demonstrates new
tools for designing markets with specific features.
\end{abstract}

\section{Basic Construction}

The simplest CSMM and CPMM curves are given by $x+y=k_{1}$ and $xy=k_{2}$;
here one thinks of $x$ and $y$ as being the amounts of the different
currencies in the market. For simplicity, let's assume that $k_{1}=2$
and $k_{2}=1$ so that both curves contain the point $(x_{0},y_{0})=(1,1)$.
Note that this means that the CSMM can be rewritten as $\frac{x+y}{2}=1$;
the advantage here is that now both curves are defined with a constant
of $1$ on the righthand side and taking combinations of the two is
very easy.

For reasons that will be apparent soon, denote $A_{0}(x,y)=\frac{x+y}{2}$
and $A_{1}(x,y)=xy$. The goal is to find a family of functions $A_{t}(x,y)$
for $0\leq t\leq1$ that smoothly transitions from $A_{0}(x,y)$ to
$A_{1}(x,y)$. Perhaps the easiest way is to take an arithmetic weighted
mean of the two functions:

\begin{equation}
A_{t}^{arith}(x,y)=A_{0}(x,y)(1-t)+A_{1}(x,y)t
\end{equation}

\noindent Part of the reason this works as well as it does is because
$A_{0}(x,y)=1$ and $A_{1}(x,y)=1$; thus, any weighted average of
the two will also be equal to $1$.

A concrete example of this construction is seen in Stableswap (\cite{Ego19}).
In an $n$-currency Stableswap market the initial amounts $X^{i}=(x_{1}^{i},...,x_{n}^{i})$
of each currency are given by $x_{j}^{i}=\frac{D}{n}$ for some value
of $D$ and each value of $j$. The sum and product of these values
are then given by the following:

\[
\begin{array}{ccc}
\sum_{j=1}^{n}x_{j}^{i}=D &  & \prod_{j=1}^{n}x_{j}^{i}=\left(\frac{D}{n}\right)^{n}\end{array}
\]

\noindent The Stableswap system has a self-proclaimed ``leverage''
parameter $\chi$ where $\chi=0$ corresponds to CPMM and $\chi=\infty$
to CSMM; if the quantities of each currency are given by $X=(x_{1},...,x_{n})$
then the AMM is

\[
\chi D^{n-1}\sum_{j=1}^{n}x_{j}+\prod_{j=1}^{n}x_{j}=\chi D^{n}+\left(\frac{D}{n}\right)^{n}
\]

\noindent The claim here is that these formulas can be represented
using the above $A_{t}^{arith}$ format. To see this, note that the
CSMM and CPMM equations can written as the following:

\[
\begin{array}{ccc}
A_{0}(X)=\frac{1}{D}\sum_{j=1}^{n}x_{j} &  & A_{1}(X)=\left(\frac{n}{D}\right)^{n}\prod_{j=1}^{n}x_{j}\end{array}
\]

\noindent Each of these equations is normalized so that the surfaces
are defined by $A_{0}(X)=1$ and $A_{1}(X)=1$. This in turn allows
the defining mixed equation to be written as

\noindent 
\begin{eqnarray*}
1 & = & \frac{\chi D^{n-1}\sum_{j=1}^{n}x_{j}+\prod_{j=1}^{n}x_{j}}{\chi D^{n}+\left(\frac{D}{n}\right)^{n}}\\
 & = & A_{0}(X)\frac{\chi D^{n}}{\chi D^{n}+\left(\frac{D}{n}\right)^{n}}+A_{1}(X)\frac{\left(\frac{D}{n}\right)^{n}}{\chi D^{n}+\left(\frac{D}{n}\right)^{n}}\\
 & = & A_{0}(X)\frac{\chi}{\chi+n^{-n}}+A_{1}(X)\frac{n^{-n}}{\chi+n^{-n}}
\end{eqnarray*}

\noindent Thus, the Stableswap equations can be reparametrized to
the arithmetic mixing format where the relation between $\chi$ and
$t$ is given by

\begin{equation}
t=\frac{n^{-n}}{\chi+n^{-n}}
\end{equation}

While the arithmetic mean is a simple way of combining the curves,
it is far from the only way. Intuitively, the curve $A_{t}^{arith}(x,y)$
is just an arithmetic weighted mean of the CSMM and CPMM curves. A
common alternative to the arithmetic mean is the geometric mean. For
example, recall that the arithmetic and geometric means of the numbers
$3$ and $5$ are $\frac{1}{2}(3+5)=4$ and $(3\cdot5)^{\frac{1}{2}}=\sqrt{15}$
respectively. One can define a geometric weighted mean of the CSMM
and CPMM curves in a similar way:

\begin{equation}
A_{t}^{geo}(x,y)=A_{0}(x,y)^{1-t}A_{1}(x,y)^{t}
\end{equation}

\noindent This can appear more complicated than the arithmetic version,
but it also offers several computational advantages. For example,
$A_{0}(x,y)=\frac{x+y}{2}$ and $A_{1}(x,y)=xy$ are homogeneous functions
of degree $1$ and $2$ respectively, i.e. $A_{0}(\lambda x,\lambda y)=\lambda A_{0}(x,y)$
and $A_{1}(\lambda x,\lambda y)=\lambda^{2}A_{0}(x,y)$. There are
known benefits to having a homogenous AMM curve (\cite{TPL22}) and
one can show that $A_{t}^{geo}(x,y)$ is homogenous for all values
of $t$ while $A_{t}^{arith}(x,y)$ is non-homogeneous.

A more general construction is needed to fully describe the pros and
cons of different mixing methods. The purpose is to find an optimal
balance of price stability, impermanent loss, and ability to provide
liquidity.

\section{General Construction: Homotopy\label{sec: General Construction}}

\begin{figure}[t]
\begin{centering}
\includegraphics[scale=0.75]{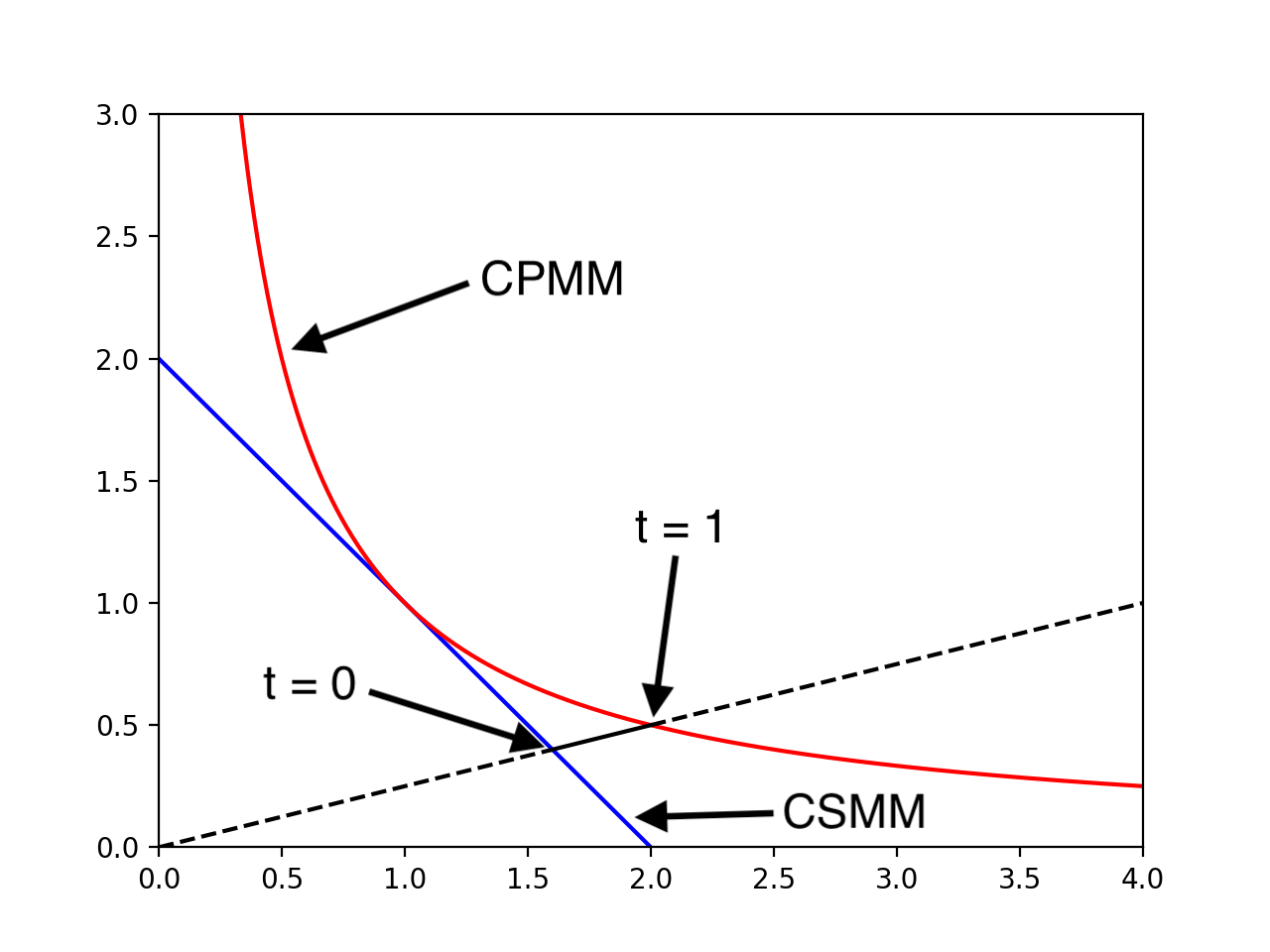}
\par\end{centering}
\caption{\label{fig: lambda figure}Visualization in 2D of the line segment
connecting the CSMM and CPMM curves. The point $(x_{0},y_{0})$ is
at the intersection of blue and red, $(\lambda_{0}x,\lambda_{0}y)$
at the intersection of blue and black, and $(\lambda_{1}x,\lambda_{1}y)$
at the intersection of red and black.}

\end{figure}

A classical mathematical construction of a ``homotopy'' between
two functions $f,g:\mathbb{R}\rightarrow\mathbb{R}^{n}$ is a weighted
average $H(s,t)=(1-t)\cdot f(s)+t\cdot g(s)$ depending on some parameter
$t$. Fixing a value of $t$ gives a very precise blending of $f$
and $g$ in that each resulting point $H(s,t)$ is exactly $100t$
percent of the way along the line segment connecting $f(s)$ and $g(s)$.
Despite their equations being visually similar, the mixings above
do not satisfy this property for a variety of reasons; values of $t$
in arithmetic and geometric mixings are only geometrically meaningful
in this way when $t$ is $0$ or $1$. The idea of the proceeding
``homotopy mixing'' is that $t$ will always represent the percent
movement from CSMM to CPMM in the above mathematical way (see Figure
\ref{fig: lambda figure} for a visual representation). Note that
the current and initial states will be denoted as $X=(x_{1},...,x_{n})$
and $X^{i}=(x_{1}^{i},...,x_{n}^{i})$ for shorthand throughout this
section.

This construction is best presented in an abstract setting where the
number of currencies is arbitrary, even though the focus of this paper
is for 2D AMMs. Suppose the CSMM and CPMM surfaces are given by

\noindent 
\begin{eqnarray*}
A_{0}(X)=\frac{\sum_{j=1}^{n}a_{j}x_{j}}{\sum_{j=1}^{n}a_{j}x_{j}^{i}}=1 &  & A_{1}(X)=\prod_{j=1}^{n}\left(\frac{x_{j}}{x_{j}^{i}}\right)^{\alpha_{j}}=1
\end{eqnarray*}

\noindent Here the $x_{j}$'s are the current quantities of the currencies
in the AMM and the $x_{j}^{i}$'s are the initial quantities. Both
of these surfaces pass through the same initial point $X^{i}$; however,
it is also important for the surfaces to share the same exchange rates
at that initial state. To find the condition for this, differentiation
shows that the gradients of these CSMM and CPMM equations are given
by

\noindent 
\begin{eqnarray*}
\nabla A_{0}(X)=\frac{1}{\sum_{j=1}^{n}a_{j}x_{j}^{i}}(a_{1},...,a_{n}) &  & \nabla A_{1}(X)=\prod_{j=1}^{n}\left(\frac{x_{j}}{x_{j}^{i}}\right)^{\alpha_{j}}\left(\frac{\alpha_{1}}{x_{1}},...,\frac{\alpha_{n}}{x_{n}}\right)
\end{eqnarray*}

\noindent One may think of the gradient as giving the prices of the
two currencies, at least up to some common multiple. One can get the
exchange rate from one currency to another by looking at the quotient
of the two corresponding gradient values. For example, the exchange
rate of exchanging the $x_{1}$ currency for the $x_{2}$ currency
is given by $\frac{a_{2}}{a_{1}}$ and $\frac{\alpha_{2}x_{1}^{i}}{\alpha_{1}x_{2}^{i}}$
in $A_{0}$ and $A_{1}$ respectively. Intuitively this makes sense,
especially for $A_{1}$; as the amount of the $x_{2}$ currency grows
its value will go down and so it costs fewer of the $x_{1}$ currency
to get a particular amount of $x_{2}$ currency. If one assumes that
the $A_{0}$ and $A_{1}$ exchange rates must be equal at the initial
point $X^{i}$ then

\begin{equation}
(a_{1},...,a_{n})\text{\thinspace\thinspace\thinspace must be parallel to\thinspace\thinspace\thinspace}\left(\frac{\alpha_{1}}{x_{1}^{i}},...,\frac{\alpha_{n}}{x_{n}^{i}}\right)
\end{equation}

\noindent Note that a very natural and completely acceptable choice
from a mathematical standpoint would be to simply take $\alpha_{j}=a_{j}x_{j}^{i}$
for each $j$. However, practically speaking it is best to normalize
these values to be much smaller in order to avoid blowups; a normalization
condition of $\sum_{j=1}^{n}\alpha_{j}=1$ helps fix this issue.

There is a very natural bijective association between the CSMM and
CPMM surfaces. More specifically, any point on the CPMM surface has
a unique corresponding point on the CSMM surface, and the same correspondence
is true in the other direction as long as none of the quantities $x_{j}$
are $0$. To construct this:
\begin{itemize}
\item Suppose $v=(v_{1},...,v_{n})$ is a point in the state space $\mathbb{R}_{>0}^{n}$,
i.e. where $v_{j}>0$ for all $j$.
\item There exists some scalar function $\lambda_{0}=\lambda_{0}(v)$ such
that the point $\lambda_{0}v=(\lambda_{0}v_{1},...,\lambda_{0}v_{n})$
lies on the CSMM surface. In particular:

\[
A_{0}(\lambda_{0}v)=1\implies\lambda_{0}=\frac{\sum_{j=1}^{n}a_{j}x_{j}^{i}}{\sum_{j=1}^{n}a_{j}v_{j}}
\]

\item Similarly, there exists some scalar function $\lambda_{1}=\lambda_{1}(v)$
such that the point $\lambda_{1}v=(\lambda_{1}v_{1},...,\lambda_{1}v_{n})$
lies on the CPMM surface. By denoting $deg(A_{1})=\sum_{j=1}^{n}\alpha_{j}$:

\[
A_{1}(\lambda_{1}v)=1\implies\lambda_{1}=\prod_{j=1}^{n}\left(\frac{x_{j}^{i}}{v_{j}}\right)^{\frac{\alpha_{j}}{deg(A_{1})}}
\]

\end{itemize}
The correspondence is then given by associating $\lambda_{0}v$ with
$\lambda_{1}v$. This method creates a pairing between the surfaces
as a whole because one could always choose $v$ to be an arbitrary
point on the CSMM or the CPMM.

This correspondence between CSMM and CPMM state points allows for
a natural description of a smooth transition from one to the other.
This method is inspired by the notion of homotopy in algebraic topology,
and therefore the constructed surface will be denoted $A_{t}^{hom}(x_{1},...,x_{n})=1$.
This function is constructed as follows. Suppose $x=(x_{1},...,x_{n})$
is a point on this homotopy surface for the given parameter $t$.
The assumption in this construction is that the point $x$ lies $100t$
percent of the way along the line segment connecting $\lambda_{0}x$
to $\lambda_{1}x$. The equation form of this assumption is given
by

\noindent 
\begin{eqnarray*}
x & = & (1-t)\lambda_{0}x+t\lambda_{1}x\\
 & = & ((1-t)\lambda_{0}+t\lambda_{1})x
\end{eqnarray*}

\noindent Clearly if $t=0$ then $x=\lambda_{0}x$ and so $x$ must
lie on the CSMM surface; similarly, if $t=1$ then $x=\lambda_{1}x$
and so $x$ is on the CPMM surface. Note that the 2D version of this
argument is presented in Figure \ref{fig: lambda figure}.

More generally, the only way the above equation $x=((1-t)\lambda_{0}+t\lambda_{1})x$
holds is if $(1-t)\lambda_{0}+t\lambda_{1}=1$ (because all values
$x_{1},...,x_{n}$ are assumed strictly positive). This equation is
exactly the needed one as original the goal was to find a function
that was equal to $1$ that described points on the surface. Upon
simplifying, the result becomes

\noindent 
\begin{eqnarray*}
A_{t}^{hom}(X) & = & (1-t)\frac{\sum_{j=1}^{n}a_{j}x_{j}^{i}}{\sum_{j=1}^{n}a_{j}x_{j}}+t\prod_{j=1}^{n}\left(\frac{x_{j}^{i}}{x_{j}}\right)^{\frac{\alpha_{j}}{deg(A_{1})}}\\
 & = & A_{0}(X)^{-1}(1-t)+A_{1}(X)^{-\frac{1}{deg(A_{1})}}t
\end{eqnarray*}

The remainder of this section is concerned with this specifics of
this construction in 2D. The parameter $t$ along with the addition
of another parameter $s$ allows for a complete parametric description
of the region between the CSMM and CPMM curves in 2D. As seen in Section
\ref{sec: Advanced-Construction}, such a parametrization is a very
powerful tool for designing and building mixing methods that satisfy
any range of desired properties.

Before deriving any equations using the parameter $s$, denote the
following for simplicity moving forward:

\noindent 
\begin{eqnarray*}
A_{0}(x,y)=\frac{ax+by}{ax_{0}+by_{0}} &  & A_{1}(x,y)=\frac{x^{\alpha}y^{\beta}}{x_{0}^{\alpha}y_{0}^{\beta}}
\end{eqnarray*}

\noindent The above is a $t$-parametric description of the line segment
connecting the pair of points on the CSMM and CPMM surfaces; one can
describe the original CSMM curve given by $A_{0}(x,y)=1$ in a similar
parametric way. The $y$-intercept and $x$-intercept of this line
are given by $(0,\frac{a}{b}x_{0}+y_{0})$ and $(x_{0}+\frac{b}{a}y_{0},0)$
respectively. As above, one can connect these two points using a line
segment parametrized by the variable $s$. The CSMM position corresponding
to the value of $s$ is the point

\[
(1-s)\left(0,\frac{a}{b}x_{0}+y_{0}\right)+s\left(x_{0}+\frac{b}{a}y_{0},0\right)\text{\thinspace\thinspace\thinspace or equivalently\thinspace\thinspace\thinspace}\left(ax_{0}+by_{0}\right)\left(\frac{s}{a},\frac{1-s}{b}\right)
\]

\noindent One may think of $s$ as determining the slope of the dotted
line in Figure \ref{fig: lambda figure}, or equivalently the angle
that the dotted line makes with the x-axis. Next, suppose $(x,y)$
is some point on this curve, i.e. $A_{t}^{hom}(x,y)=1$. Given a fixed
point $(x,y)$ on the curve $A_{t}^{hom}$, one can solve for the
$s$ corresponding to $(x,y)$ using that fact that $(x,y)$ and $(\frac{s}{a},\frac{1-s}{b})$
lie on the same line passing through the origin. In particular, the
ratios of the two coordinates in each point must be the same:

\[
\frac{y}{x}=\frac{\frac{1-s}{b}}{\frac{s}{a}}
\]

\begin{equation}
\implies s=\frac{ax}{ax+by}
\end{equation}

\section{Comparison of Arithmetic, Geometric, and Homotopy Mixings}

Section \ref{sec: General Construction} demonstrates a rigorous construction
of a method for mixing CSMM and CPMM curves that is more grounded
in measuring the literal distance travelled from one curve to the
another than previous methods have been. This concreteness comes with
the trade-off of having a potentially more complicated defining equation.
As such, it is important to directly compare the three above mixing
methods to see and analyze all the pros and cons. The most relevant
properties are as follows: existence of a convenient parametrization,
ability to provide liquidity, whether or not the curve is exchange
rate level independent, and the amount of stability in exchange rates
given changes in currency quantity.

\begin{table}[t]

\begin{centering}
\begin{tabular}{|>{\centering}p{1.5in}|>{\centering}p{4.2in}|}
\hline 
Mixing Type & Equation in 2D\tabularnewline
\hline 
\hline 
$\begin{array}{c}
\\
\\
\\
\end{array}$Arithmetic$\begin{array}{c}
\\
\\
\\
\end{array}$ & $A_{t}^{arith}(x,y)=\left(\frac{ax+by}{ax_{0}+by_{0}}\right)(1-t)+\left(\frac{x^{\alpha}y^{\beta}}{x_{0}^{\alpha}y_{0}^{\beta}}\right)t$\tabularnewline
\hline 
$\begin{array}{c}
\\
\\
\\
\end{array}$Geometric$\begin{array}{c}
\\
\\
\\
\end{array}$ & $A_{t}^{geo}(x,y)=\left(\frac{ax+by}{ax_{0}+by_{0}}\right)^{1-t}\left(\frac{x^{\alpha}y^{\beta}}{x_{0}^{\alpha}y_{0}^{\beta}}\right)^{t}$\tabularnewline
\hline 
$\begin{array}{c}
\\
\\
\\
\end{array}$Homotopy$\begin{array}{c}
\\
\\
\\
\end{array}$ & $A_{t}^{hom}(x,y)=\left(\frac{ax_{0}+by_{0}}{ax+by}\right)(1-t)+\left(\frac{x_{0}^{\alpha}y_{0}^{\beta}}{x^{\alpha}y^{\beta}}\right)^{\frac{1}{\alpha+\beta}}t$\tabularnewline
\hline 
\end{tabular}
\par\end{centering}
\medskip{}

\medskip{}

\medskip{}

\begin{centering}
\begin{tabular}{|>{\centering}p{1.5in}|>{\centering}p{4.2in}|}
\hline 
Mixing Type & Equation in General\tabularnewline
\hline 
\hline 
$\begin{array}{c}
\\
\\
\\
\end{array}$Arithmetic$\begin{array}{c}
\\
\\
\\
\end{array}$ & $A_{t}^{arith}(x_{1},...,x_{n})=A_{0}(x_{1},...,x_{n})(1-t)+A_{1}(x_{1},...,x_{n})t$\tabularnewline
\hline 
$\begin{array}{c}
\\
\\
\\
\end{array}$Geometric$\begin{array}{c}
\\
\\
\\
\end{array}$ & $A_{t}^{geo}(x_{1},...,x_{n})=A_{0}(x_{1},...,x_{n})^{1-t}A_{1}(x_{1},...,x_{n})^{t}$\tabularnewline
\hline 
$\begin{array}{c}
\\
\\
\\
\end{array}$Homotopy$\begin{array}{c}
\\
\\
\\
\end{array}$ & $A_{t}^{hom}(x_{1},...,x_{n})=A_{0}(x_{1},...,x_{n})^{-1}(1-t)+A_{1}(x_{1},...,x_{n})^{-\frac{1}{deg(A_{1})}}t$\tabularnewline
\hline 
\end{tabular}
\par\end{centering}
\caption{\label{tab: mixing type table}The three main types of CSMM and CPMM
mixings considered in this paper. The AMM is always defined by $A_{t}(x,y)=1$.
The parameter $t$ controls the transition from CSMM at $t=0$ to
CPMM at $t=1$. The top table gives the equations in this specific
2D context; the bottom table provides a more abstract presentation
distanced from the specifics of the CSMM and CPMM surfaces in higher
dimensions.}

\end{table}

\subsection{Parametrization}

The above parametric homotopy construction allows for the easy plotting
of the homotopy mixing curves $A_{t}^{hom}(x,y)=1$. The point $(\frac{s}{a},\frac{1-s}{b})$
can be thought of as the base parametrization of a line segment. As
seen in Section \ref{sec: General Construction}, multiplying this
by $ax_{0}+by_{0}$ moves that point to the CSMM curve. Similarly,
multiplying by $((\frac{ax_{0}}{s})^{\alpha}(\frac{by_{0}}{1-s}))^{\frac{1}{\alpha+\beta}}$
places the point on the CPMM curve. Thus, one may parametrize the
scalar function needed to move this point to $A_{t}^{hom}$ in the
following way:

\begin{equation}
\lambda^{hom}(s,t)=(ax_{0}+by_{0})(1-t)+\left(\left(\frac{ax_{0}}{s}\right)^{\alpha}\left(\frac{by_{0}}{1-s}\right)^{\beta}\right)^{\frac{1}{\alpha+\beta}}t
\end{equation}

\noindent This equation provides a convenient closed form equation
for moving points into the homotopy mixing curve. One can show that
a similar equation can be used to move points onto the geometric mixing
curve $A_{t}^{geo}$:

\[
\begin{array}{cccc}
A_{t}^{geo}(x,y)= & \left(\frac{ax+by}{ax_{0}+by_{0}}\right)^{1-t}\left(\frac{x^{\alpha}y^{\beta}}{x_{0}^{\alpha}y_{0}^{\beta}}\right)^{t}=1 &  & (x,y)=\lambda^{geo}(s,t)\left(\frac{s}{a},\frac{1-s}{b}\right)\end{array}
\]

\begin{equation}
\implies\lambda^{geo}(s,t)=(ax_{0}+by_{0})^{\frac{1-t}{(1-t)+(\alpha+\beta)t}}\left(\left(\frac{ax_{0}}{s}\right)^{\alpha}\left(\frac{by_{0}}{1-s}\right)^{\beta}\right)^{\frac{t}{(1-t)+(\alpha+\beta)t}}
\end{equation}

In contrast, it is generally not possible find such a function with
the arithmetic curve. The problem, as seen below, is that it is not
possible to directly solve for $\lambda^{arith}(s,t)$ because doing
so amounts to solving a polynomial of arbitrary and possibly non-integer
degree. While this can certainly be approximated, doing so would be
slower and less accurate than having an explicit closed-form solution.
The polynomial in question is seen below:

\[
\begin{array}{cccc}
A_{t}^{arith}(x,y)= & \left(\frac{ax+by}{ax_{0}+by_{0}}\right)(1-t)+\left(\frac{x^{\alpha}y^{\beta}}{x_{0}^{\alpha}y_{0}^{\beta}}\right)t=1 &  & (x,y)=\lambda^{arith}(s,t)\left(\frac{s}{a},\frac{1-s}{b}\right)\end{array}
\]

\begin{eqnarray*}
\implies\lambda^{arith}(s,t)\frac{1-t}{ax_{0}+by_{0}}+\lambda^{arith}(s,t)^{\alpha+\beta}\left(\frac{s}{ax_{0}}\right)^{\alpha}\left(\frac{1-s}{by_{0}}\right)^{\beta}t & = & 1
\end{eqnarray*}

In short, the homotopy and geometric mixings have very convenient
parametric descriptions while the arithmetic mixing does not. There
are several benefits to having such a function $\lambda(s,t)$. Perhaps
the most important reason is that it allows for easy computation of
quantity changes. Keep in mind that the primary purpose of this work
is to construct an automated market maker where users can exchange
currencies. A fundamental part of such a task is providing the user
with an estimate of the total cost of exchanging for a given amount
of the desired. This is non-trivial because the AMM is defined implicitly,
so efficient computation is key here.

\begin{figure}[t]
\begin{centering}
\begin{tabular}{ccc}
\includegraphics[scale=0.3]{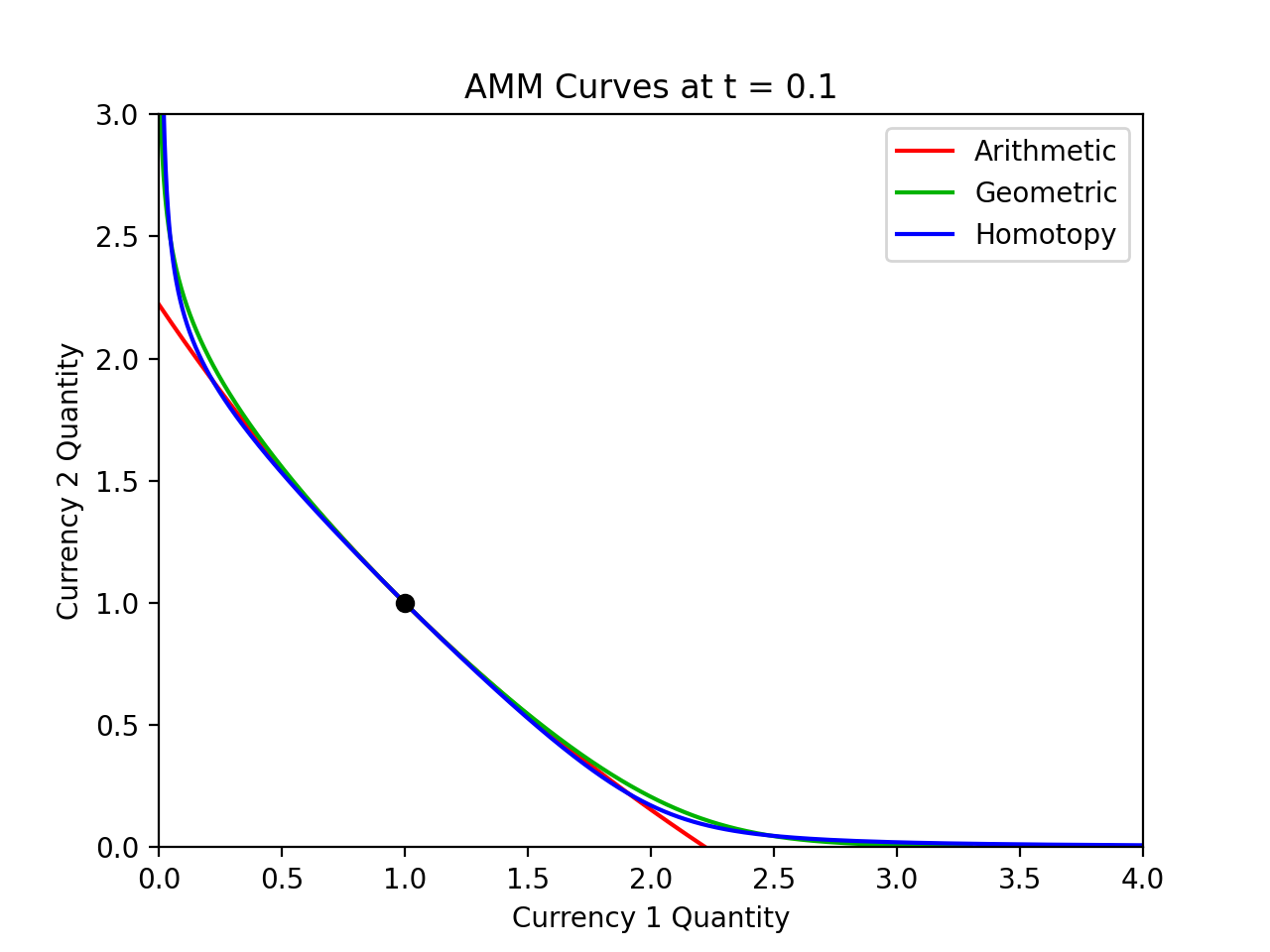} & \includegraphics[scale=0.3]{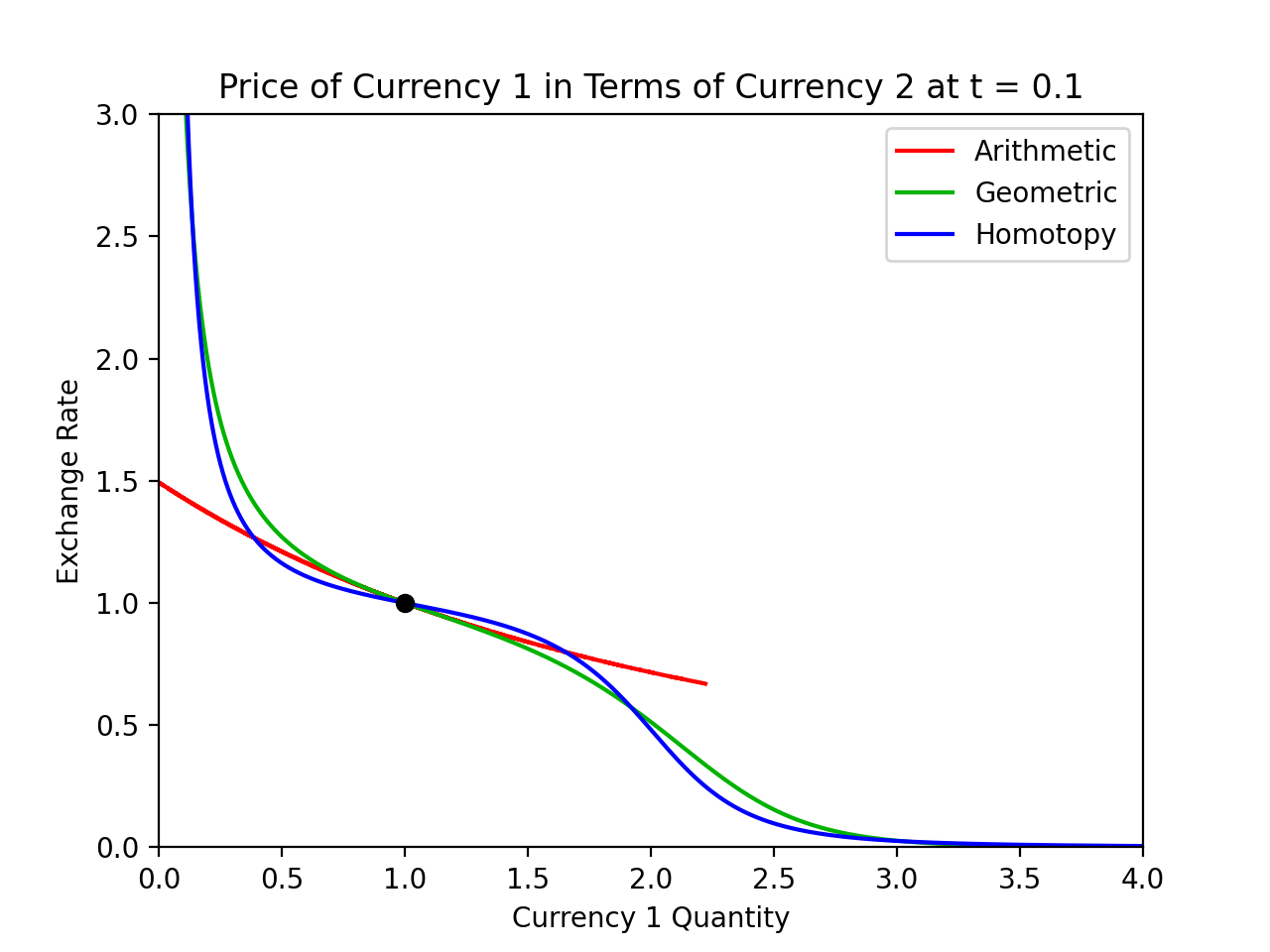} & \includegraphics[scale=0.3]{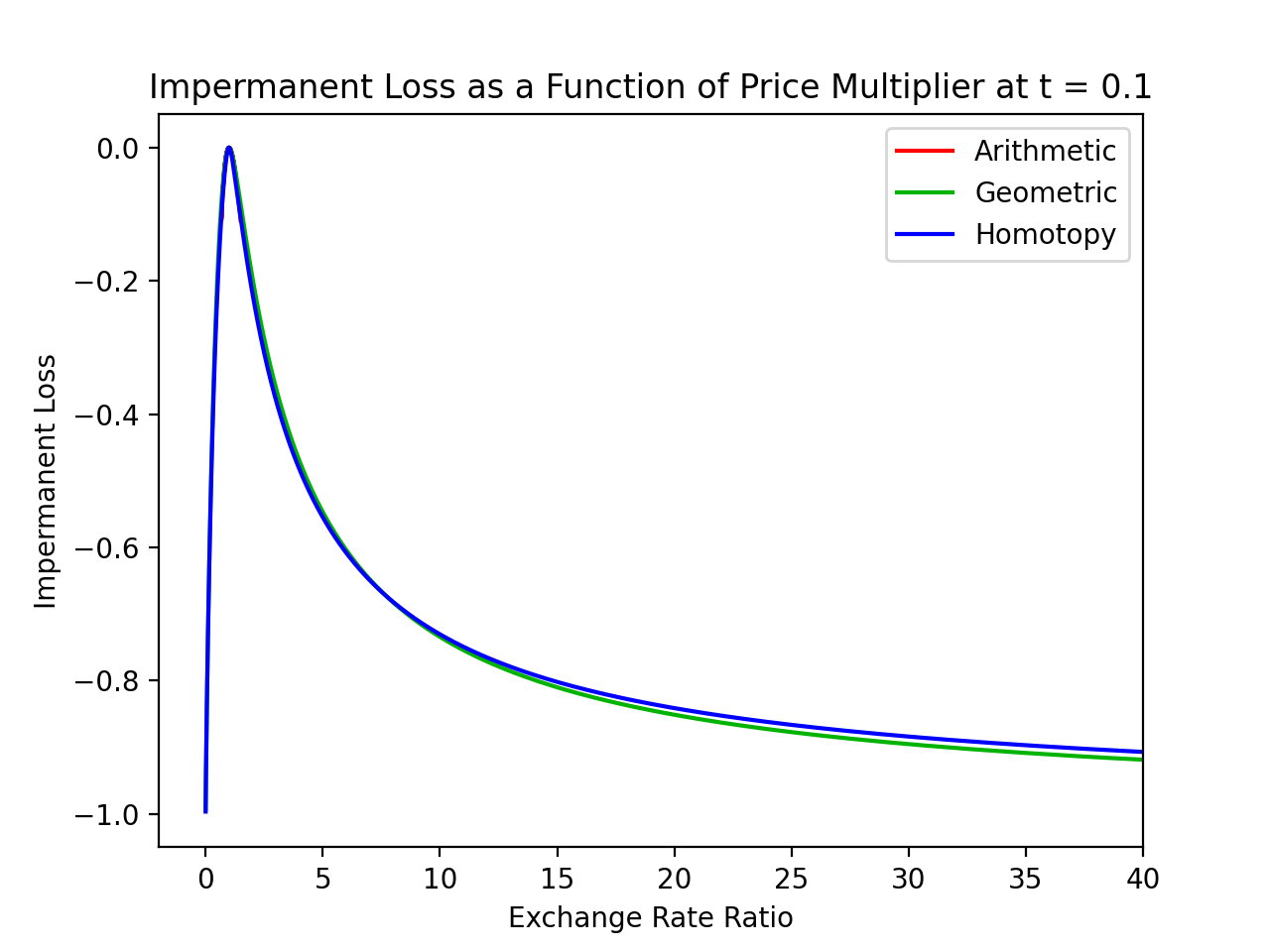}\tabularnewline
\includegraphics[scale=0.3]{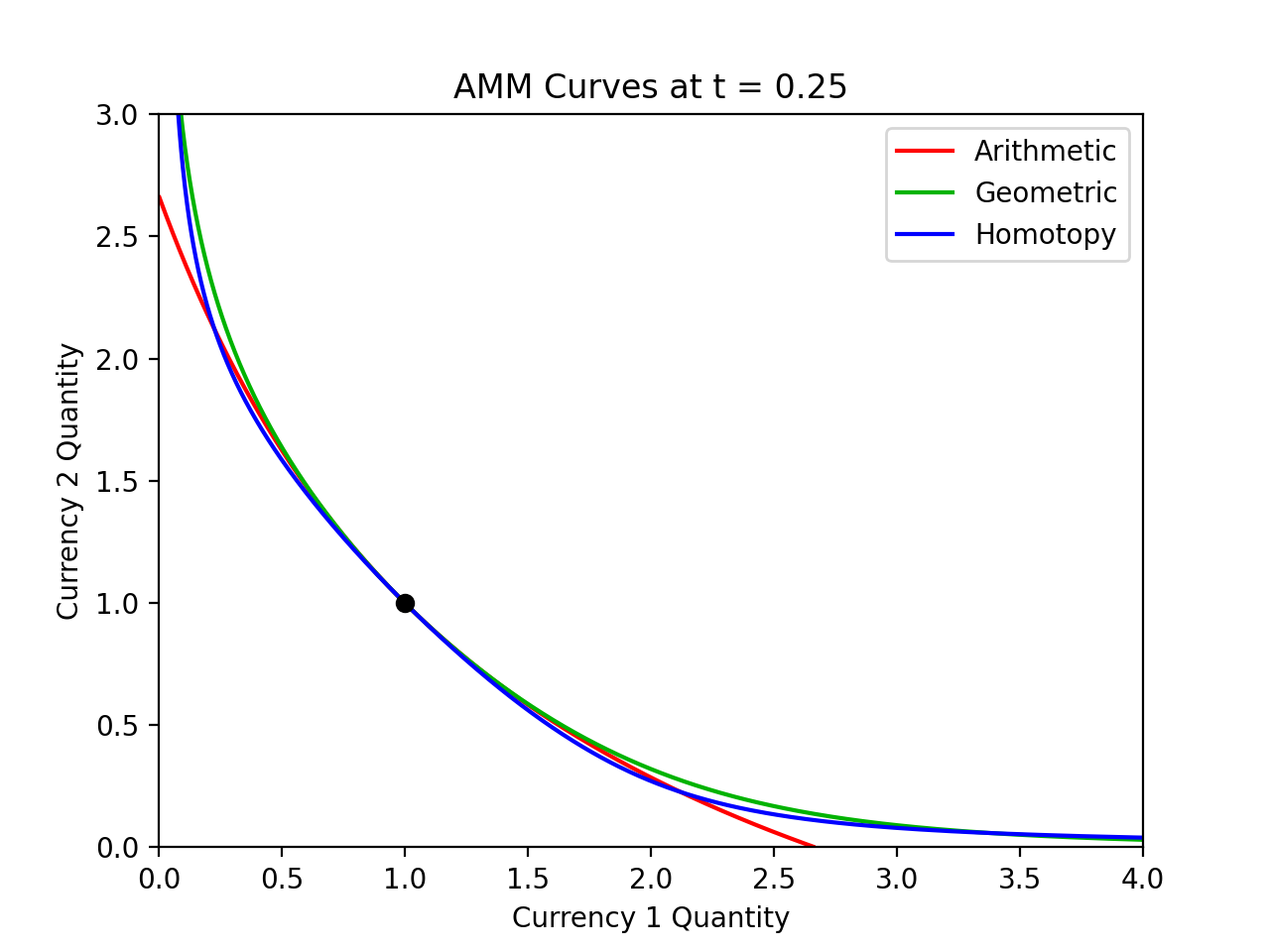} & \includegraphics[scale=0.3]{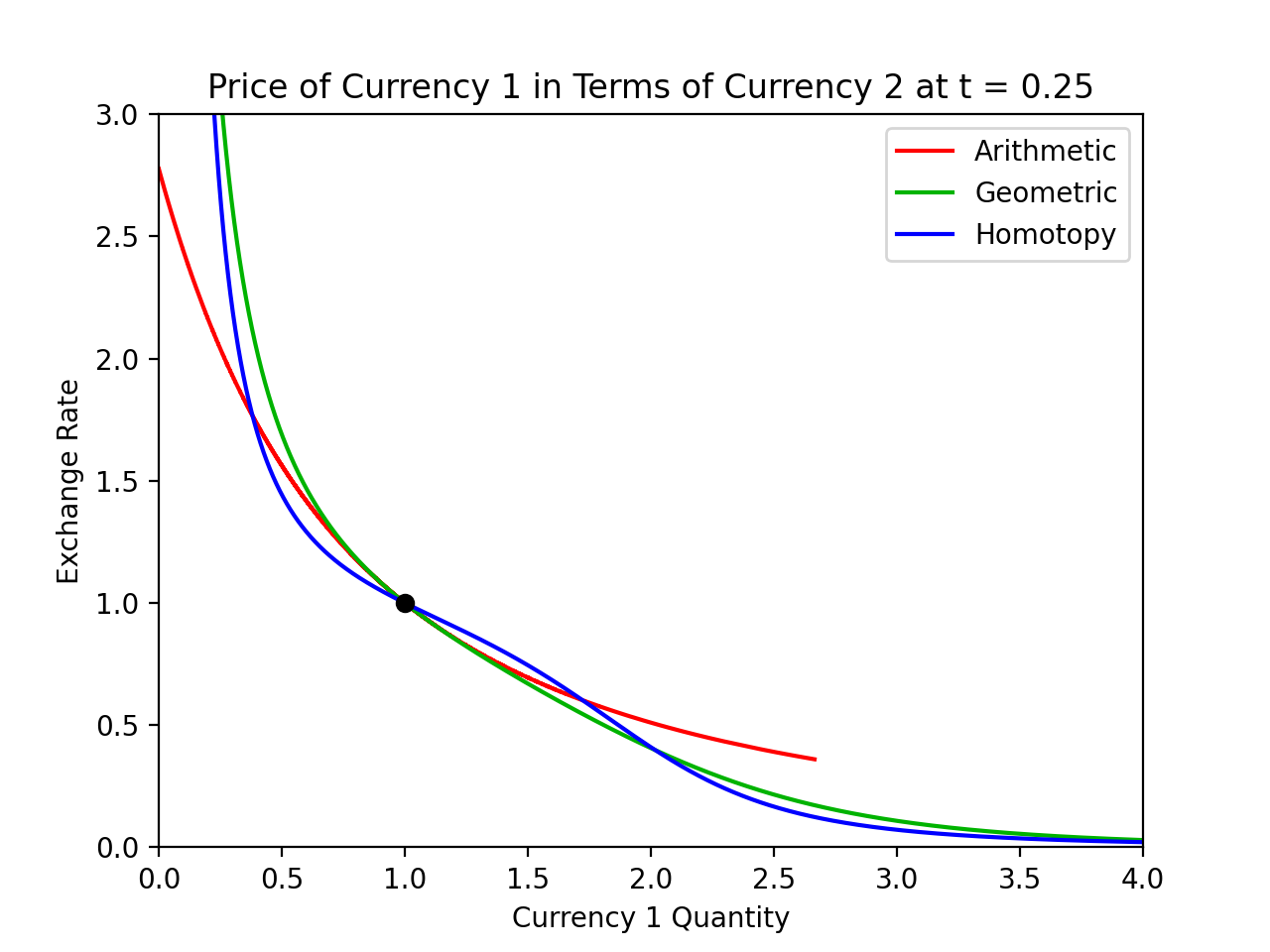} & \includegraphics[scale=0.3]{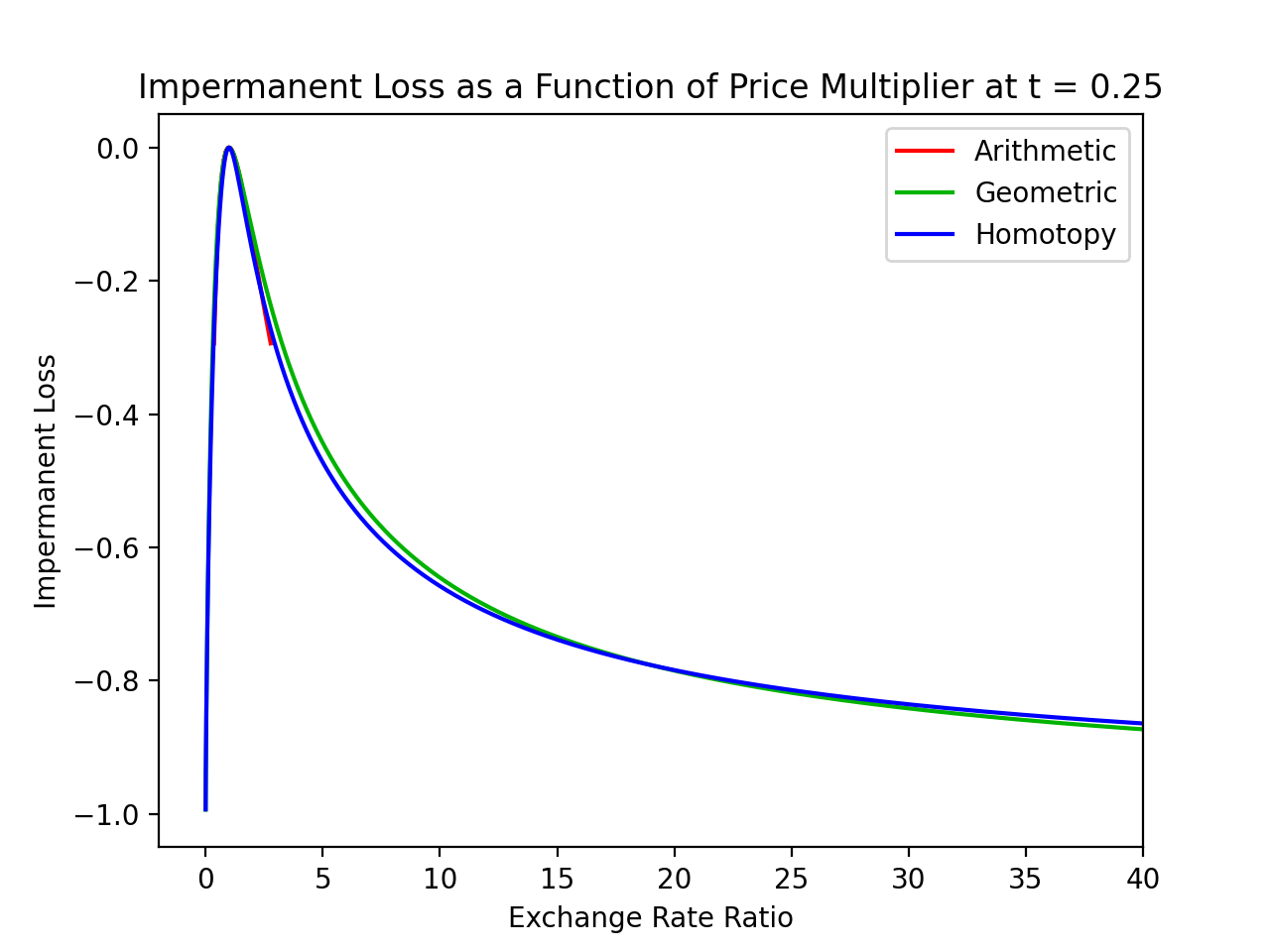}\tabularnewline
\includegraphics[scale=0.3]{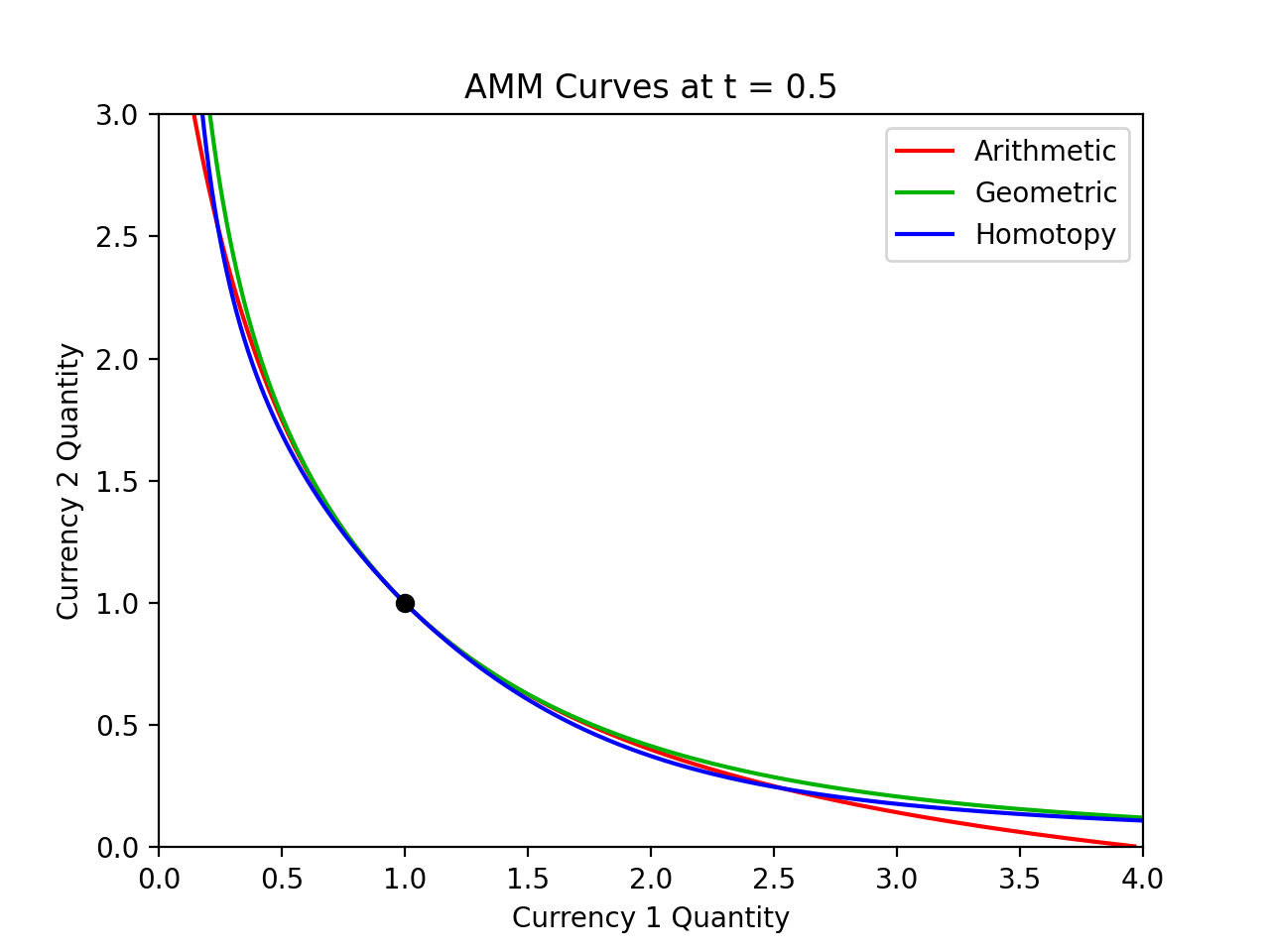} & \includegraphics[scale=0.3]{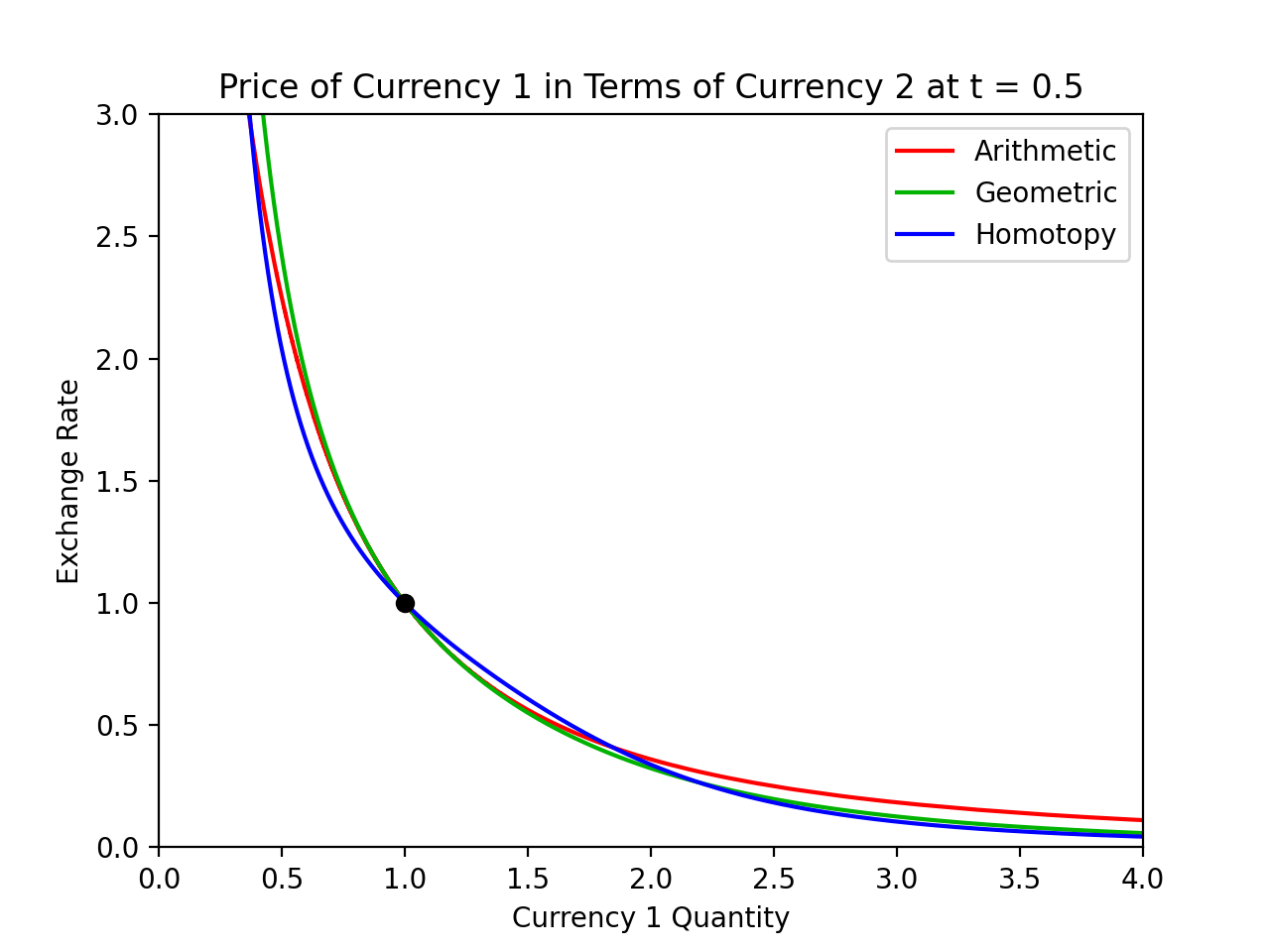} & \includegraphics[scale=0.3]{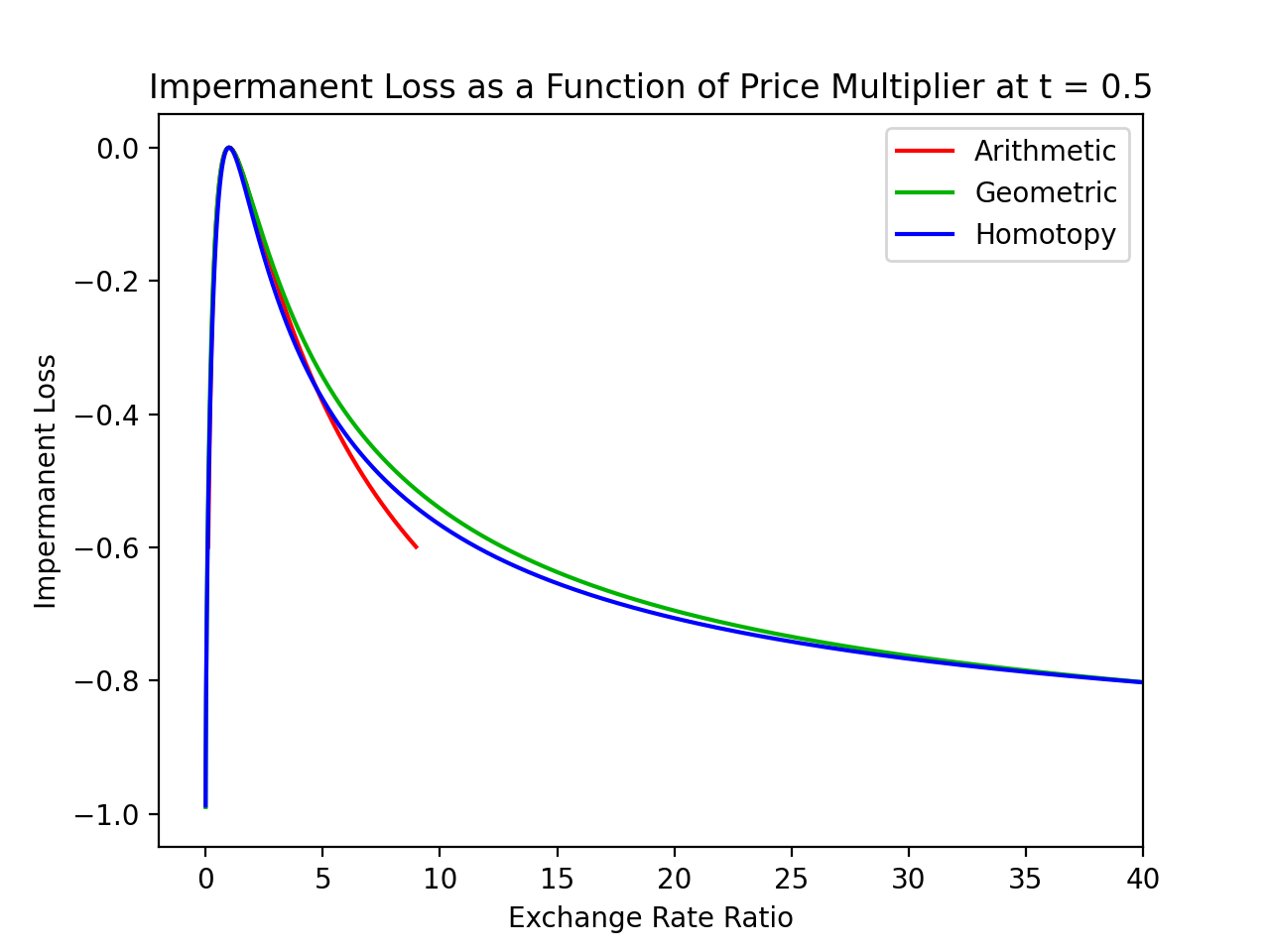}\tabularnewline
\end{tabular}
\par\end{centering}
\caption{\label{fig: curve type comparisons}Comparison of arithmetic, geometric,
and homotopy curves where each diagram has the same value of $t$
but different curve types. Important observations include: (1) arithmetic
supports a finite range of exchange rates that increases in size with
$t$, (2) geometric and homotopy support all exchange rates, and (3)
the exchange rate of homotopy the most stable of the three types for
each fixed value of $t$.}
\end{figure}

\begin{figure}[t]
\begin{centering}
\begin{tabular}{ccc}
\includegraphics[scale=0.3]{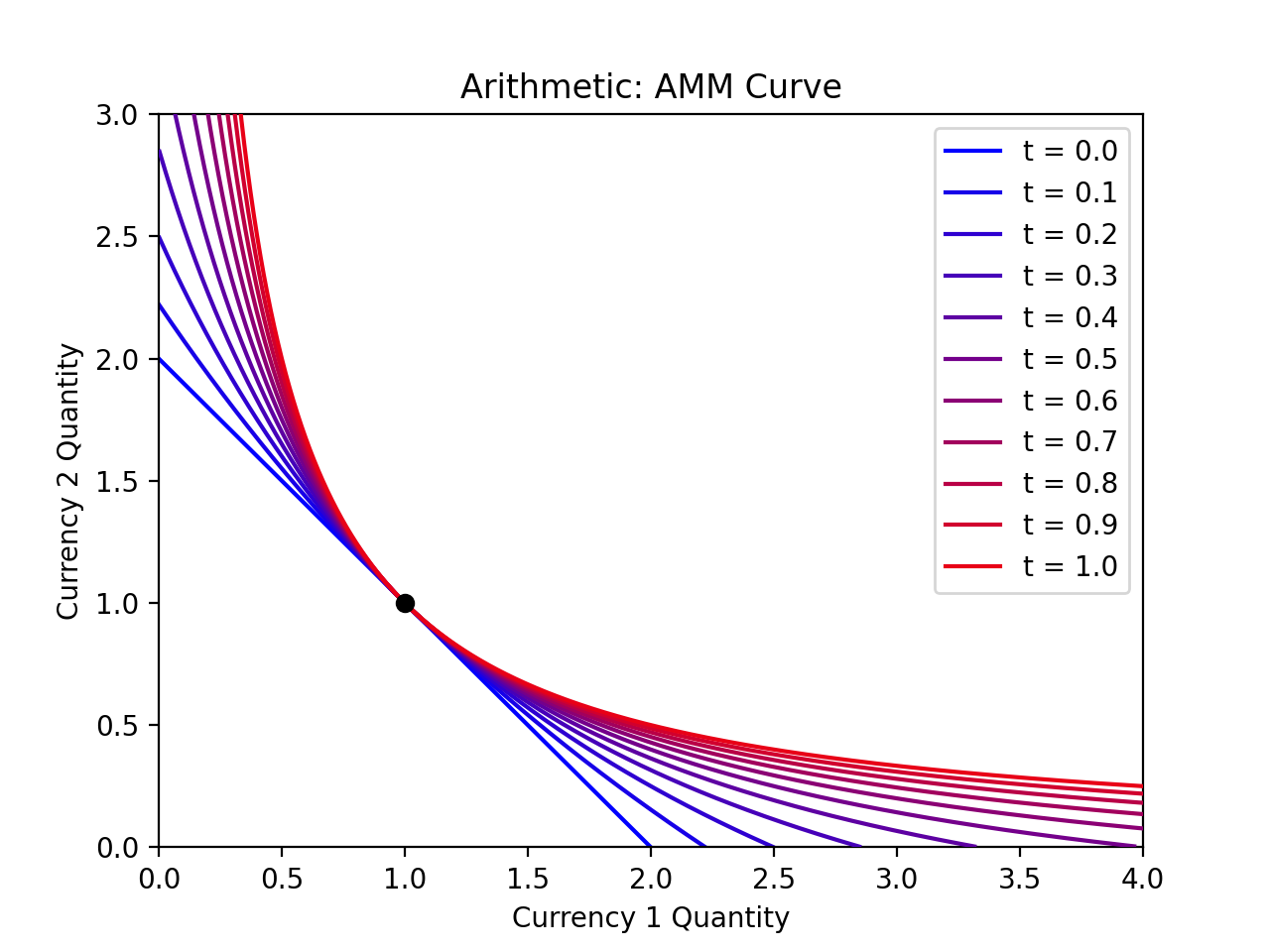} & \includegraphics[scale=0.3]{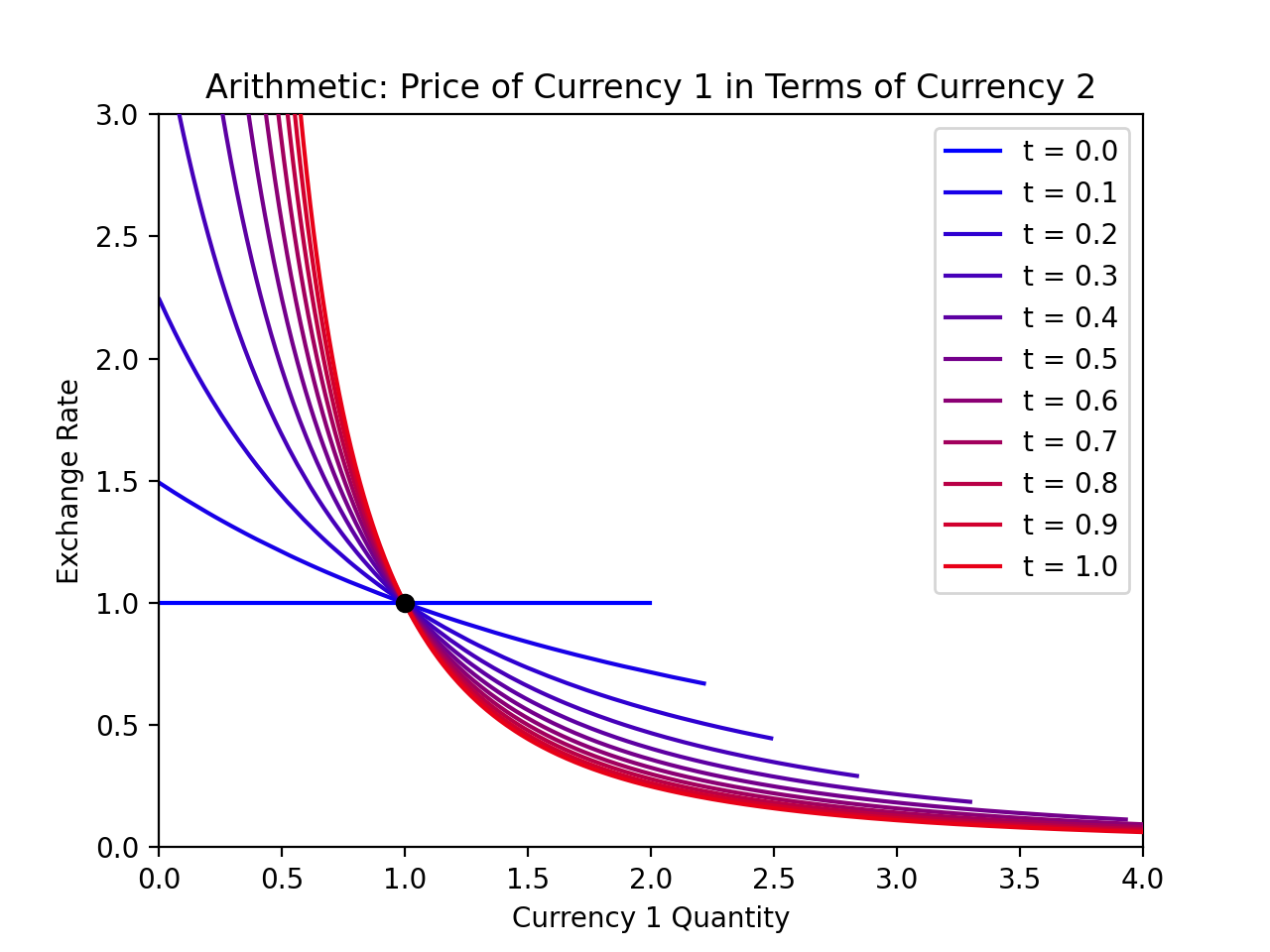} & \includegraphics[scale=0.3]{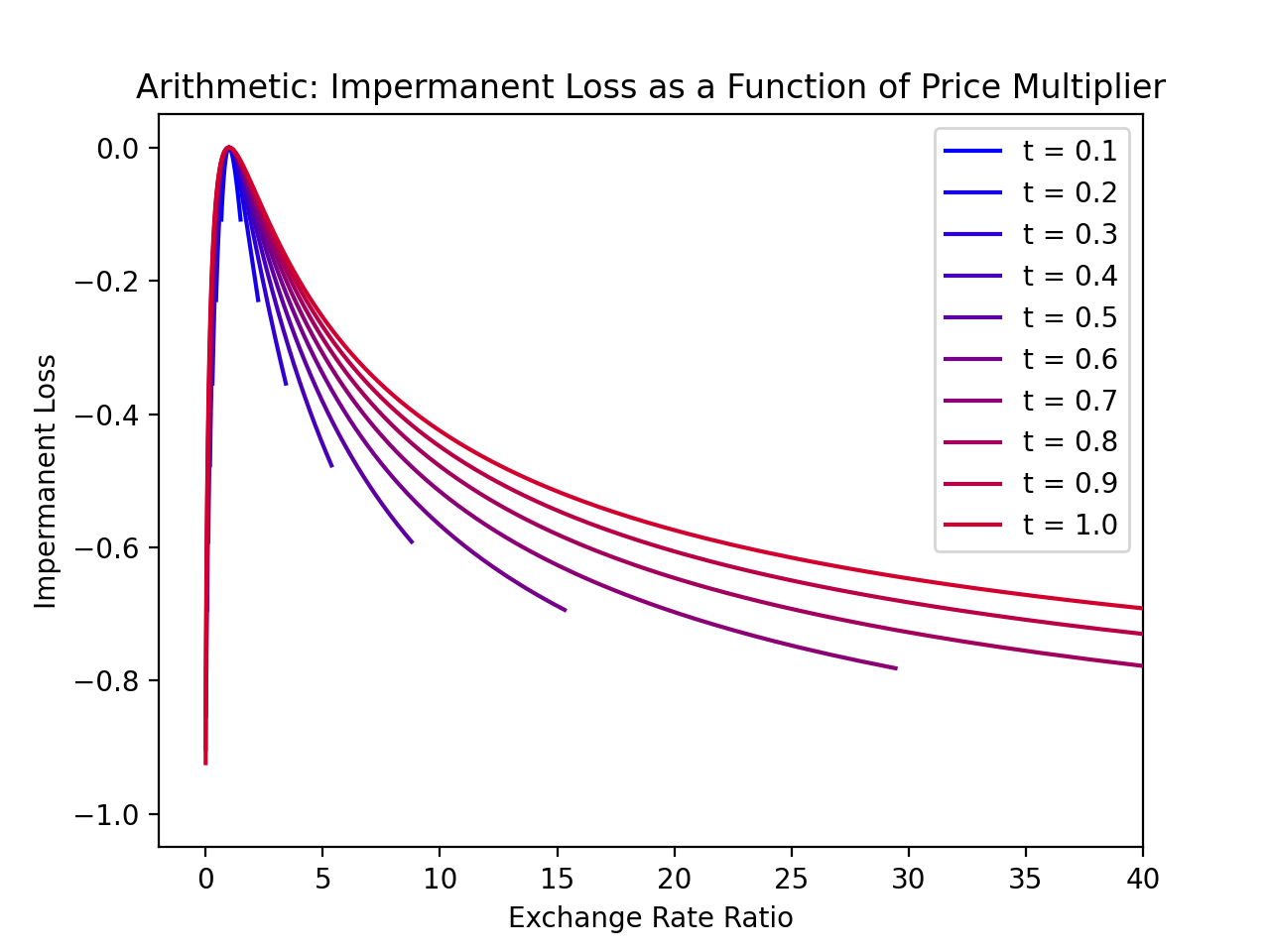}\tabularnewline
\includegraphics[scale=0.3]{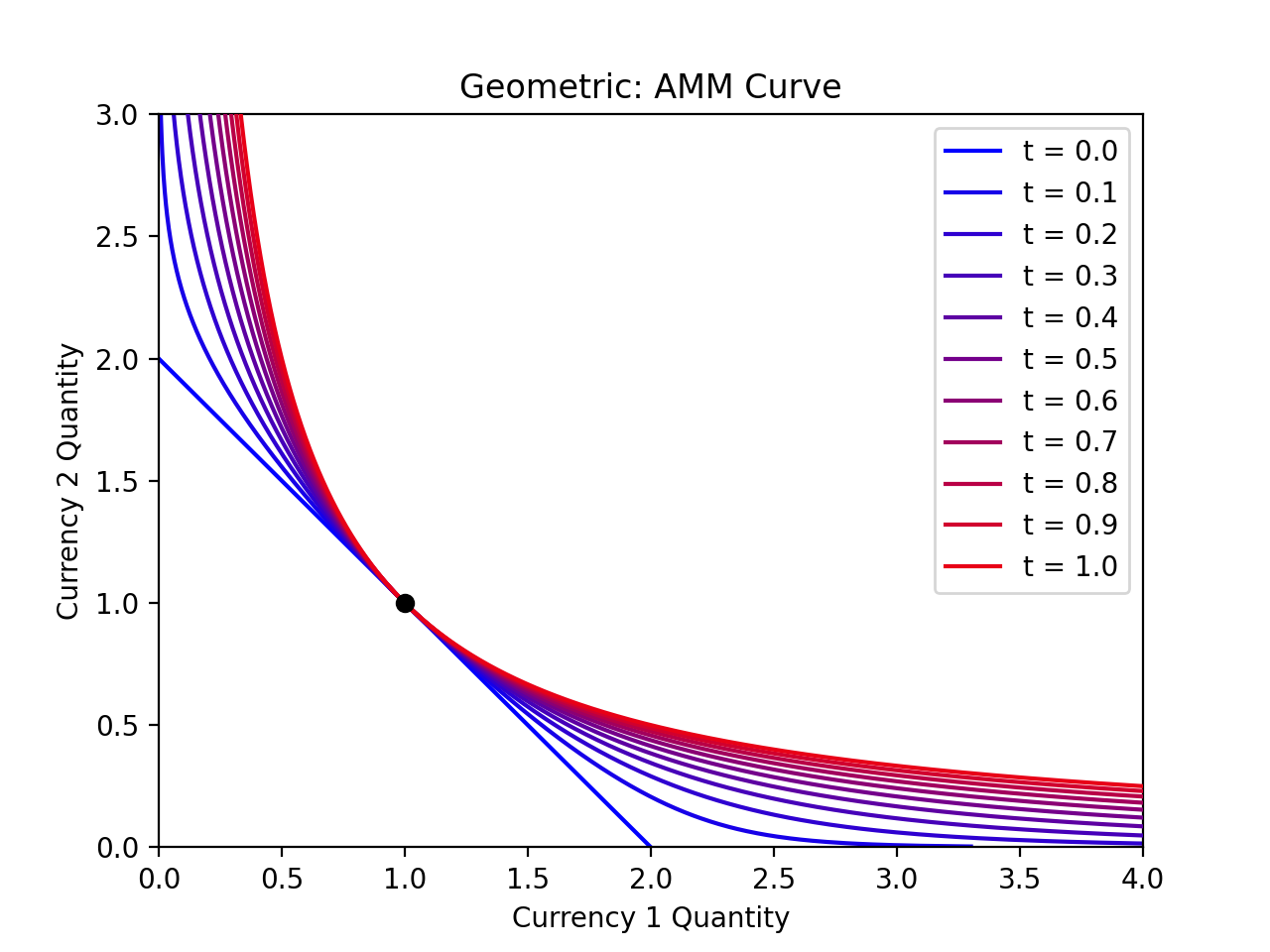} & \includegraphics[scale=0.3]{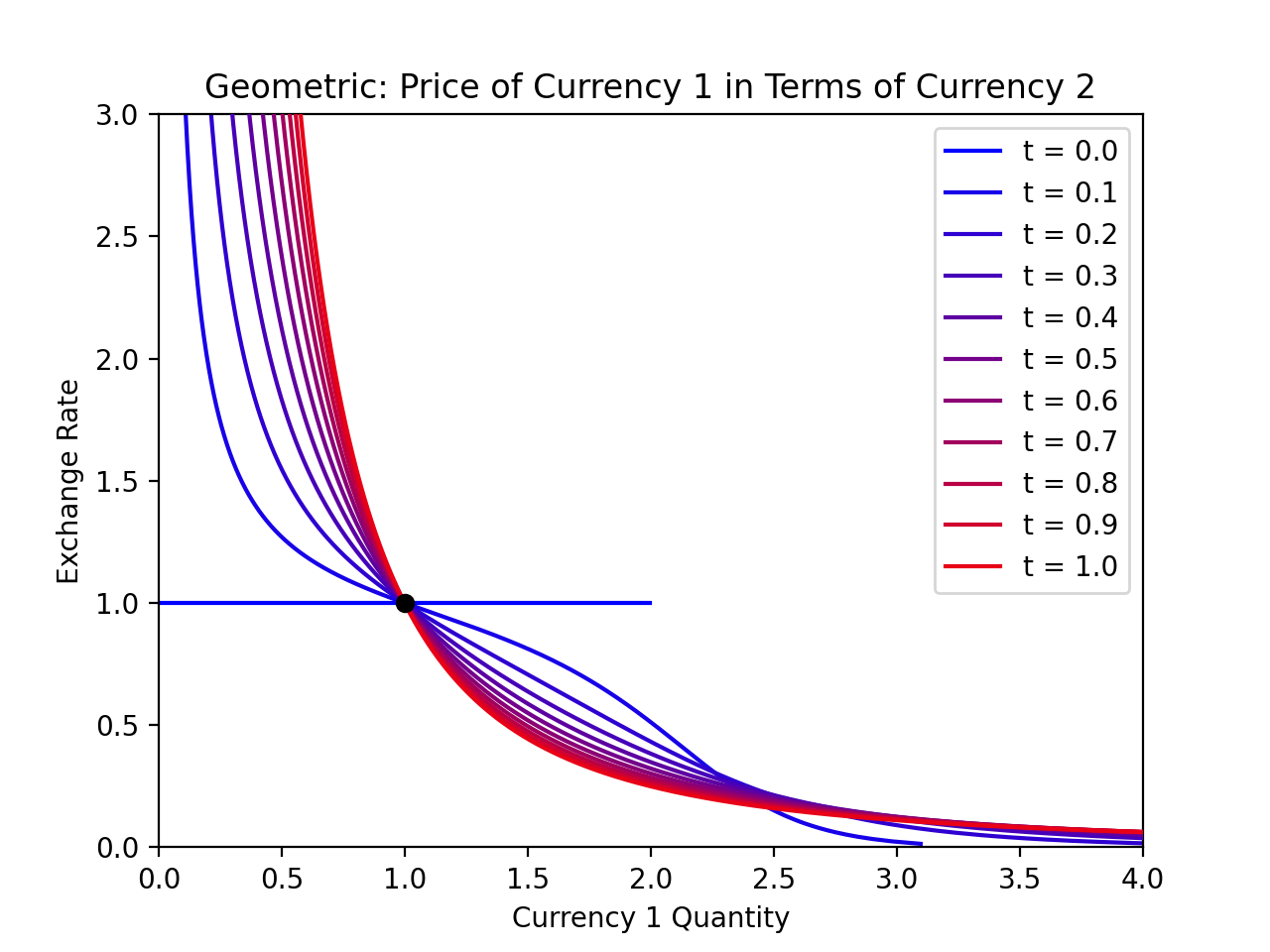} & \includegraphics[scale=0.3]{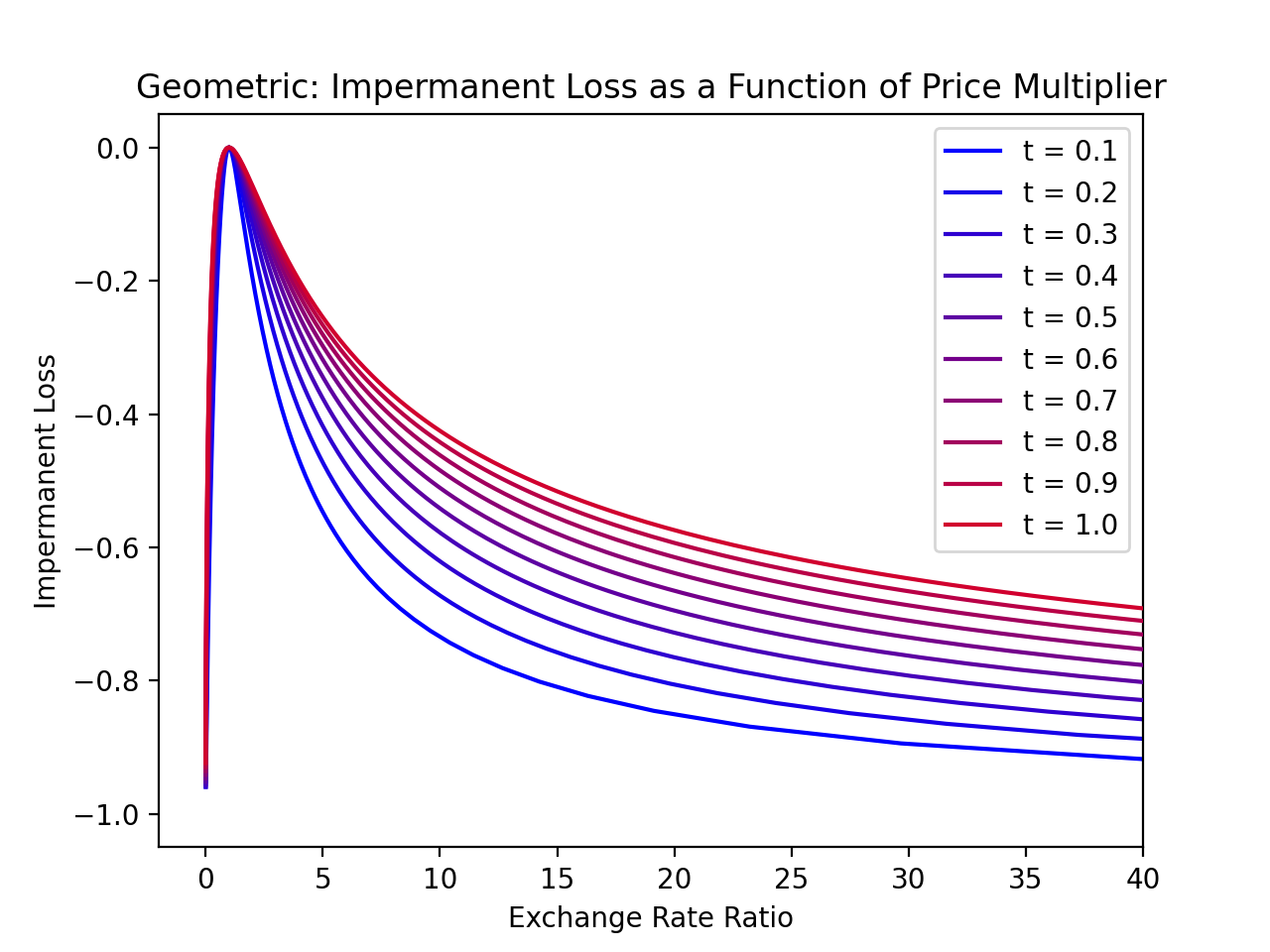}\tabularnewline
\includegraphics[scale=0.3]{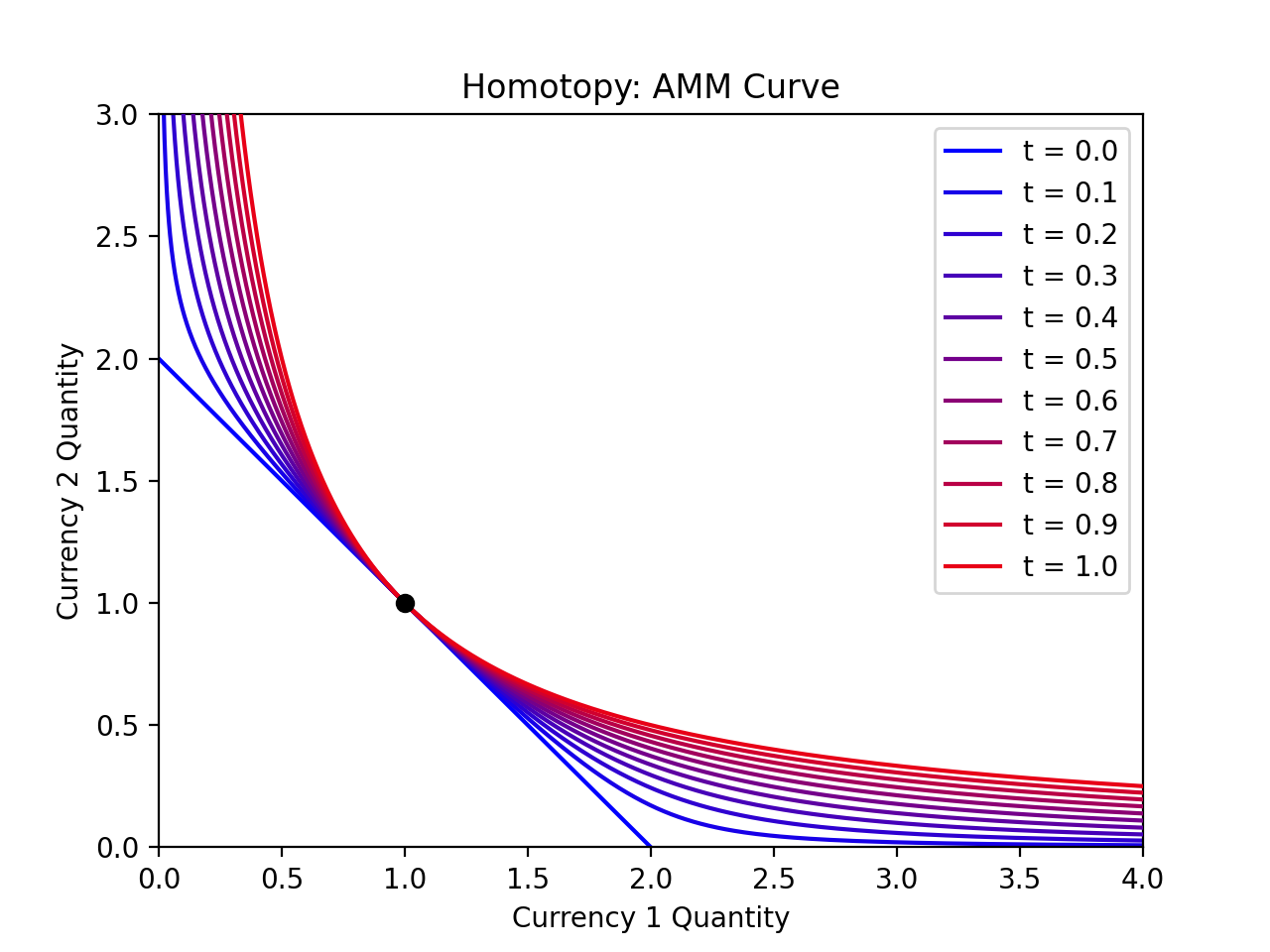} & \includegraphics[scale=0.3]{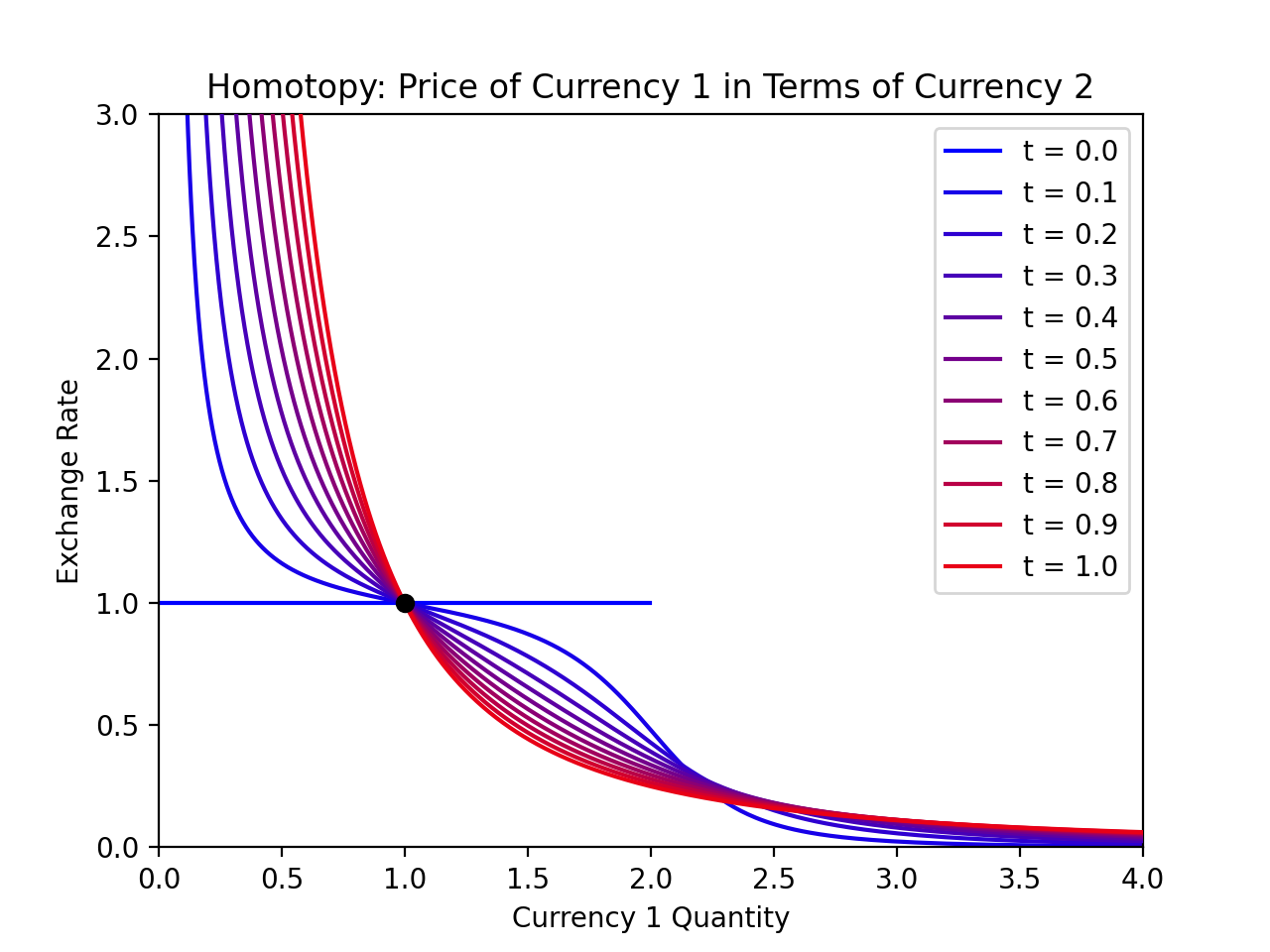} & \includegraphics[scale=0.3]{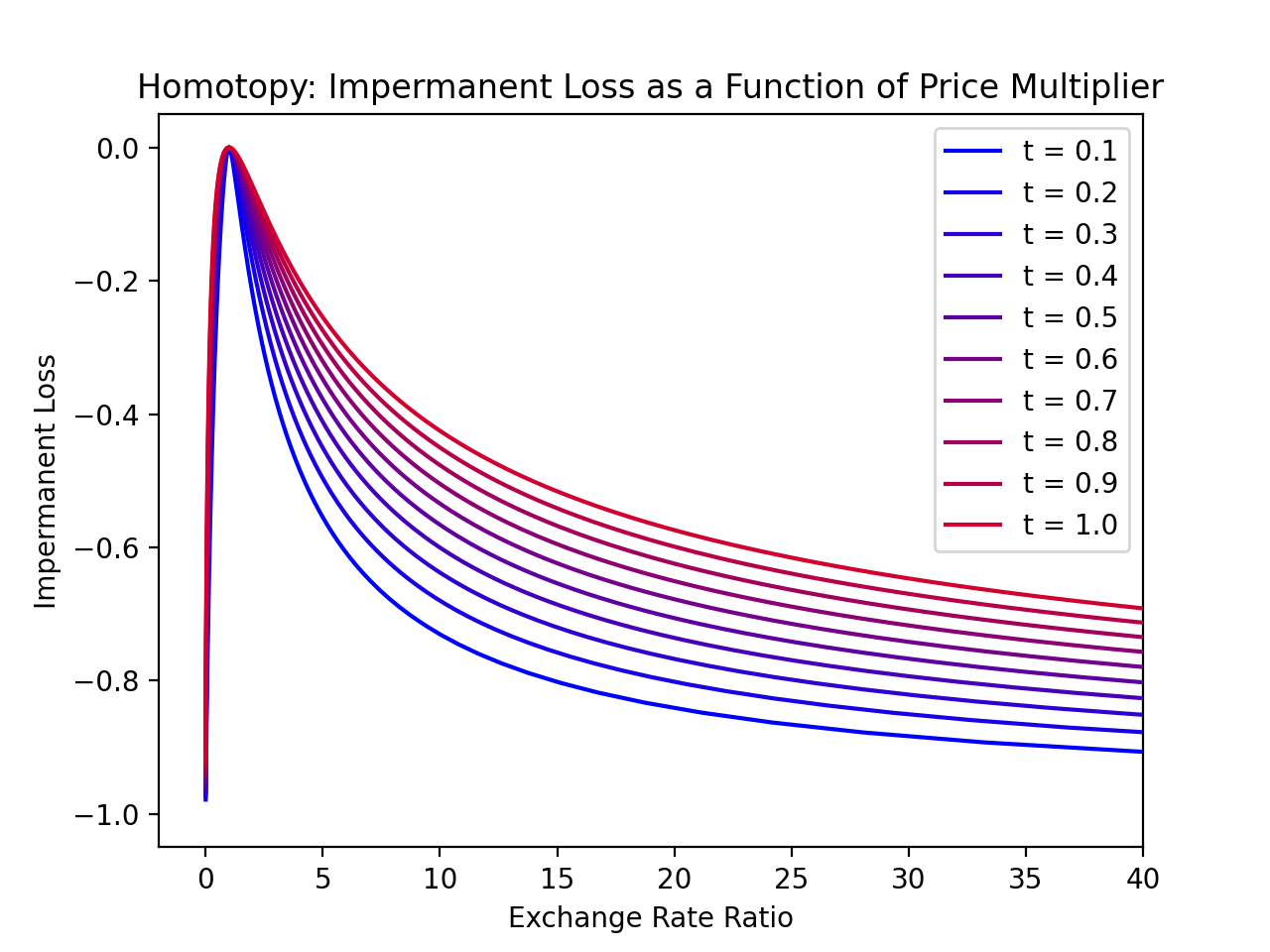}\tabularnewline
\end{tabular}
\par\end{centering}
\caption{\label{fig: curve time comparisons}Comparison of arithmetic, geometric,
and homotopy curves where each diagram is of one type at multiple
values of $t$. Important observations include: (1) price stability
decreases and impermanent loss gets worse as $t$ increases for all
three curve types, and (2) curves are more evenly distributed between
CSMM and CPMM in homotopy than they are for geometric.}

\end{figure}

\subsection{Ability to Provide Liquidity}

The CSMM and CPMM curves have several important differences, but perhaps
the most significant is the range of exchange rates that each is able
to support. One of the main defining characteristics of a CPMM is
its ability to provide liquidity at all possible exchange rates. In
such a market, as the quantity of an item goes down to zero it becomes
asymptotically more valuable; this item will never be fully extracted
because its exchange rate gets arbitrarily large. In contrast, a CSMM
only supports a single exchange rate; one can see this by the fact
that the slope of the CSMM curve is constant. As one might expect,
these mixed curves will exhibit behavior that is somewhere in between
these two extremes.

Another benefit of this $\lambda(s,t)$ function is that one can examine
the asymptotic behavior near the endpoints of mixed curves; such asymptotes
are related to the derivatives of the curve and therefore the range
of exchange rates supported by the AMM. For example, when $s=0$ or
$1$ and $t=0$ then the corresponding points are the endpoints of
the CSMM line segment; these points are finitely far away from the
origin therefore $lim_{s\rightarrow0^{+}}\lambda(s,0)$ and $lim_{s\rightarrow1^{-}}\lambda(s,0)$
must be finite for all three types of $\lambda$ functions. On the
other hand, these same limits for $\lambda(s,1)$ must approach infinity
because the CPMM ``endpoints'' are infinitely far away due to being
the asymptotes of a hyperbola.

It is important to examine the behavior of these limits when $0<t<1$. Thankfully, the asymptotes in Figures \ref{fig: curve type comparisons} and \ref{fig: curve time comparisons}
make this more clear and show that the following limits hold:

\[
\begin{array}{ccccc}
lim_{s\rightarrow0^{+}}\lambda^{arith}(s,t)<\infty &  & lim_{s\rightarrow0^{+}}\lambda^{geo}(s,t)=\infty &  & lim_{s\rightarrow0^{+}}\lambda^{hom}(s,t)=\infty\\
lim_{s\rightarrow1^{-}}\lambda^{arith}(s,t)<\infty &  & lim_{s\rightarrow1^{-}}\lambda^{geo}(s,t)=\infty &  & lim_{s\rightarrow1^{-}}\lambda^{hom}(s,t)=\infty
\end{array}
\]

\noindent One might expect that the limits in the homotopy case would be infinite because
$\lambda^{hom}$ will describe a weighted average between a finite
value and infinity; the other cases are less intuitive without the figures.

In short, the above limits show that arithmetic mixings
share the CSMM property of not always being able to provide liquidity,
and they also show that geometric and homotopy mixings more closely
resemble a CPMM in this respect. Thus, these latter two mixings are better able to match a wide range of exchange rates and this affects how arbitrageurs interact with these markets.

\subsection{Exchange Rate Level Independence for Impermanent Loss}

The idea of impermanent loss is that liquidity providers for AMMs
can be disincentivized from investing due to changes in exchange rates.
Suppose the initial quantities of assets provided to the AMM by the
investor are $X^{i}=(x_{1}^{i},...,x_{n}^{i})$ and their initial
prices are $P^{i}=(p_{1}^{i},...,p_{n}^{i})$. After a period of time,
the investor may want to withdraw their percent ownership of the AMM
and reclaim their assets. Due to changes in the market state, they
will be able to obtain quantities $X^{f}=(x_{1}^{f},...,x_{n}^{f})$
of each currency and their respective prices would be $P^{f}=(p_{1}^{f},...,p_{n}^{f})$.
The initial value of their assets is given by $P^{i}\cdot X^{i}$
and the final value of their assets is $P^{f}\cdot X^{f}$. However,
one can also consider the value they would have had if they'd simply
held onto their assets instead of investing; in that case, their held
asset value would be $P^{f}\cdot X^{i}$. The reader can refer to
\cite{TPL22} for more details, but the equation for impermanent loss
is as follows:

\begin{equation}
IL=\frac{P^{f}\cdot X^{f}}{P^{f}\cdot X^{i}}-1
\end{equation}

\noindent Roughly speaking, impermanent loss compares the relative
values of investing and holding assets. One can show that the impermanent
loss is always less than or equal to $0$ and becomes more negative
as prices drift further from initial values. In other words, the investor
has no reason to provide liquidity for the AMM unless they benefit
from collection of transaction fees that cancel out the difference
in these two values.

Figures \ref{fig: curve type comparisons} and \ref{fig: curve time comparisons}
demonstrate the impermanent loss curves for the mixed AMMs. There
are several aspects of these figures that are worth noting. First,
all curves demonstrate the property that impermanent loss is worse
the more the exchange rates change from the initial state at the time
of investment. Second, the curves closer to CPMM with larger values
of $t$ have less severe impermanent loss than those close to CSMM.
Third, it is common practice to describe impermanent loss in terms
of exchange rates even though the above formula for $IL$ is in terms
of prices and quantities. 

An AMM is ``exchange rate level independent'' (i.e. ERLI) if its
equation for impermanent loss can be written purely in terms of the
ratios of final to initial exchange rates of the currencies in the
market. The paper \cite{TPL22} gives a variety of lemmas and theorems
related to when an AMM is ERLI; note that many of these statements
relate to homogeneity. However, the application of these claims to
the equations of Table \ref{tab: mixing type table} show that none
of the three mixings satisfy this condition when $0<t<1$ and therefore
none of these mixings are ERLI.

\subsection{Slippage and Exchange Rate Stability}

Perhaps the central purpose of a platform like Stableswap is to ensure
that AMM exchange rates closely match external exchange rates when
there are small changes in AMM currency quantities. A CSMM is the
best in this respect; the exchange rate is fixed regardless of the
quantities of currencies. On the other hand, even relatively small
transactions can have sizable affects on exchange rates in a CPMM.
This price stability is relevant for users of the AMM because it helps
to minimize price slippage (i.e. the difference between the price
users expect and the actual price given by a variety of factors).

It is important to keep in mind that there is a distinction between
internal and external changes in exchange rate. Stability in internal
exchange rates must be balanced with the ability to provide liquidity
at a variety of exchange rates. For example, the CSMM is completely
stable with respect to internal changes but can only support a single
external exchange rate; if the internal and exchange exchange rates
don't match then arbitrageurs will completely drain the currency that
the AMM undervalues and thereby shut down the usefulness of the AMM.
As discussed above, a CPMM is different and can support the providing
of liquidity at any exchange rate.

\begin{figure}[t]
\begin{centering}
\begin{tabular}{cc}
\includegraphics[scale=0.5]{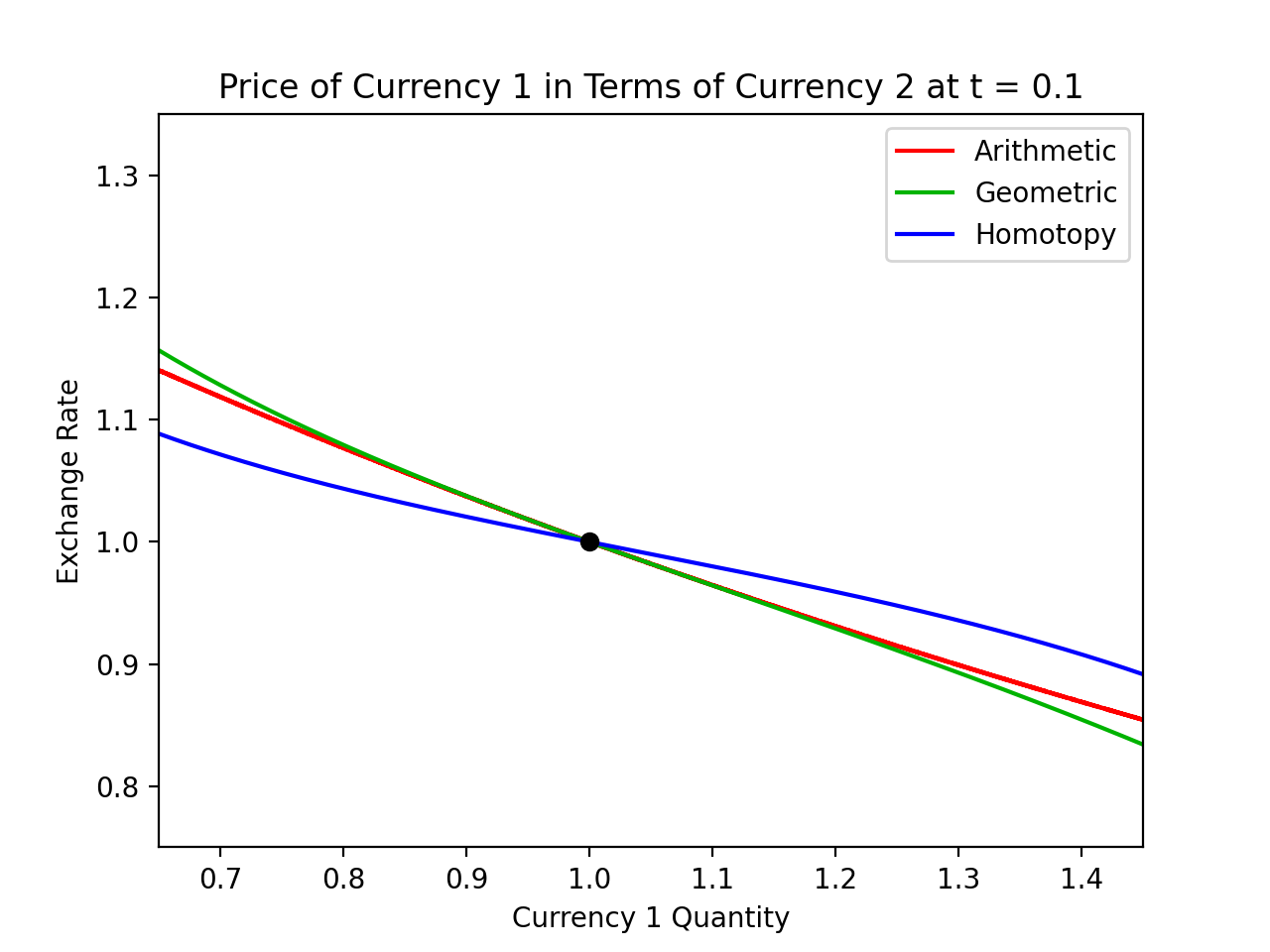} & \includegraphics[scale=0.5]{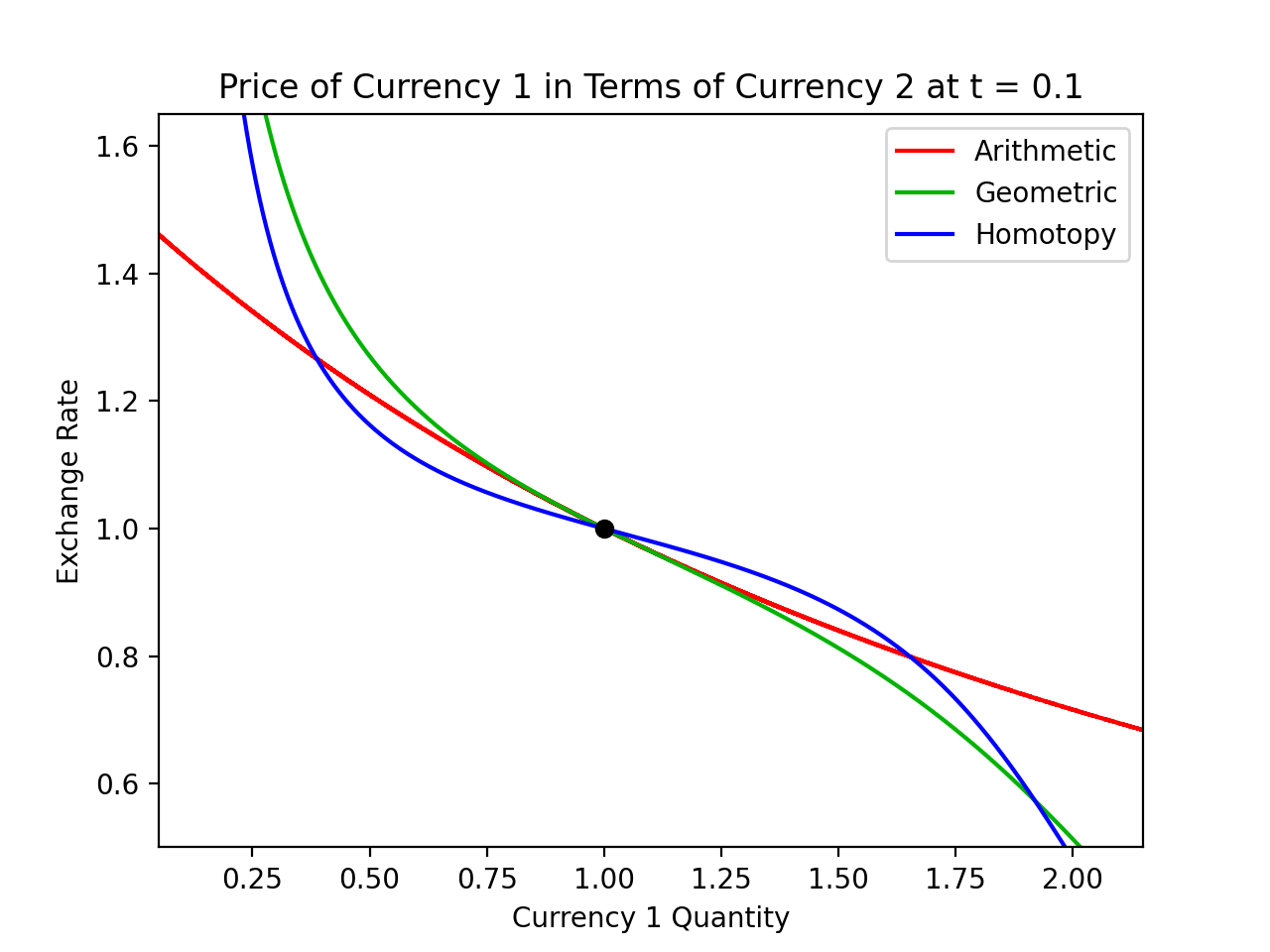}\tabularnewline
\includegraphics[scale=0.5]{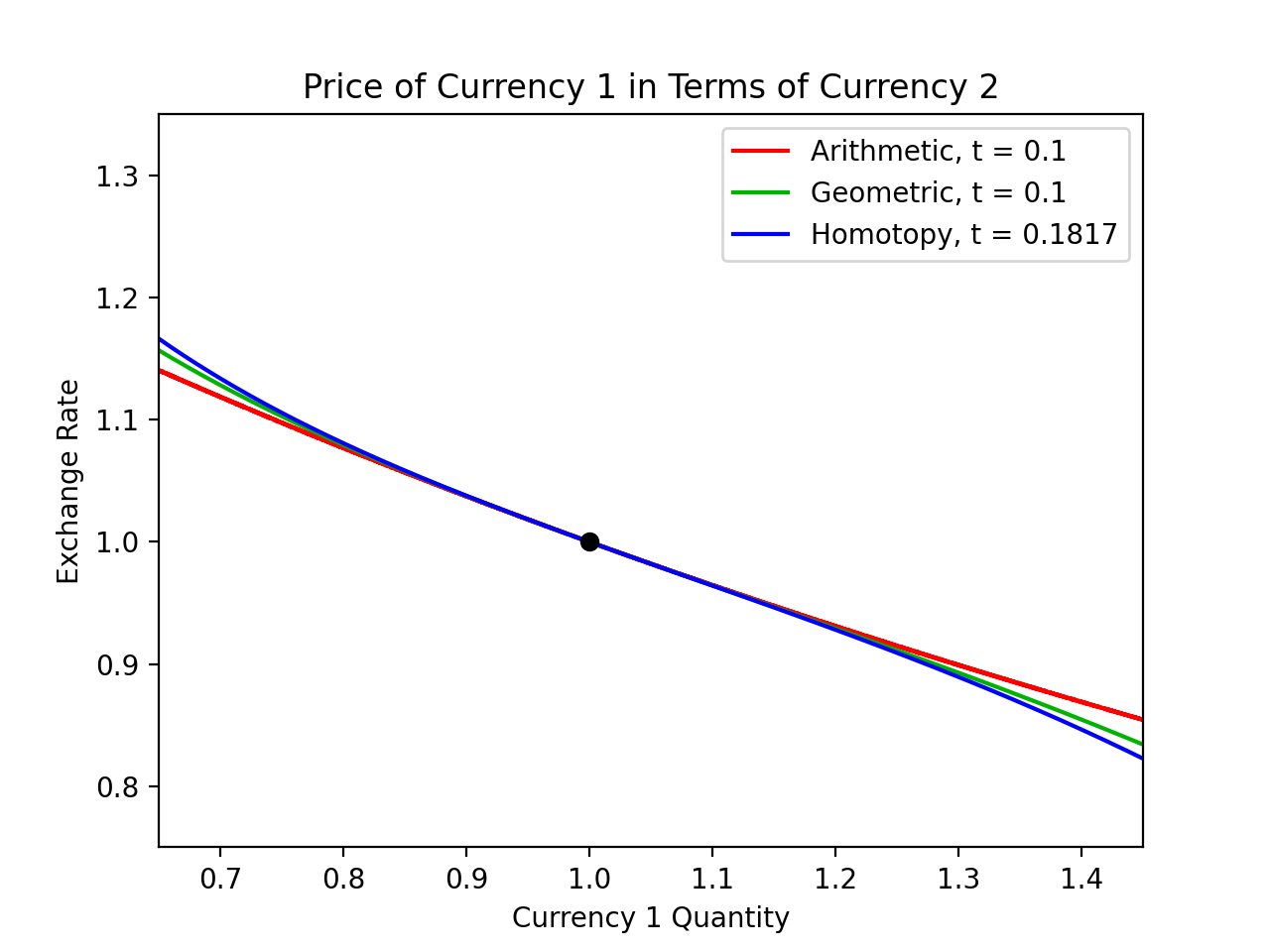} & \includegraphics[scale=0.5]{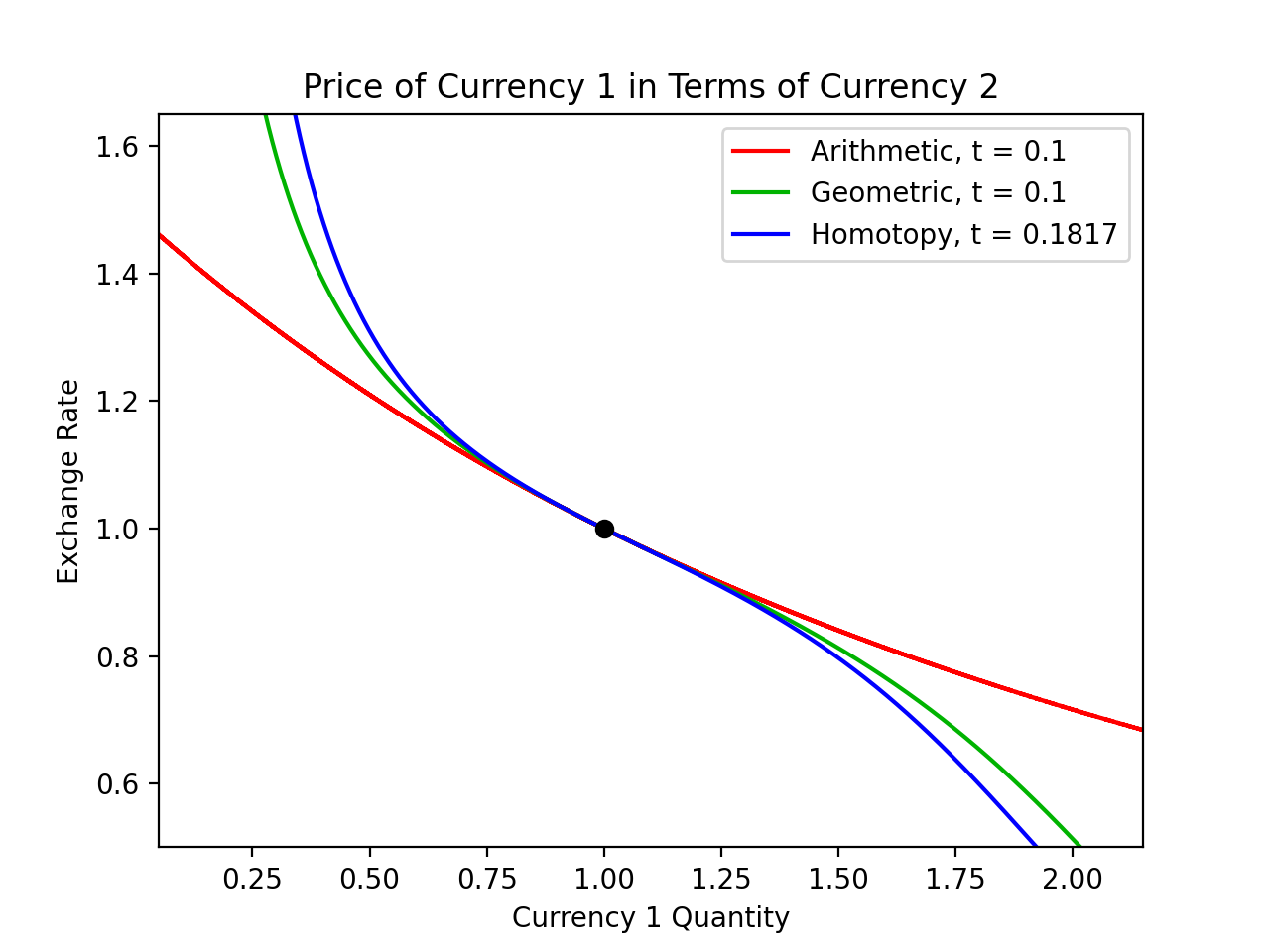}\tabularnewline
\end{tabular}
\par\end{centering}
\caption{\label{fig: stability comparison}Comparison of exchange rate stability
in the three mixed curves. (Top) These images all have the same value
of $t$. Homotopy is the most stable for small changes but the least
stable for larger ones. Geometric is the least stable for small changes.
Arithmetic is the most stable for large changes. (Bottom) These images
all have the same derivative for exchange rate at the initial point.
For both small and large changes, arithmetic is the most stable while
homotopy is the least stable. The arithmetic and geometric curves
will always have the same derivative here, but homotopy will be different.
Note that the value of $0.1817$ for $t$ was determined empirically
for this figure.}
\end{figure}

With that in mind, the diagrams in Figures \ref{fig: curve type comparisons}
and \ref{fig: curve time comparisons} show the relative stabilities
of the three mix types. For small changes in quantity at a fixed shared
value of $t$, the homotopy curve $A_{t}^{hom}(x,y)$ is the most
stable and the geometric curve $A_{t}^{geo}(x,y)$ is the least stable.
On the other hand, the arithmetic curve $A_{t}^{arith}(x,y)$ is the
most stable for large changes and the homotopy curve $A_{t}^{hom}(x,y)$
is the least. In short, each mixing type appears to have its own unique
levels of stability in different regions of the market state. These
ideas are also demonstrated in a more direct way in Figure \ref{fig: stability comparison}.

\begin{table}[t]
\begin{centering}
\begin{tabular}{|>{\centering}p{1.5in}||>{\centering}p{1.5in}|>{\centering}p{1.5in}|>{\centering}p{1.5in}|}
\hline 
 & Arithmetic & Geometric & Homotopy\tabularnewline
\hline 
\hline 
Closer to CSMM or CPMM & CSMM & CPMM & CPMM\tabularnewline
\hline 
Ability to provide liquidity & Can fail to provide liquidity & Always provides liquidity & Always provides liquidity\tabularnewline
\hline 
ERLI? & No & No & No\tabularnewline
\hline 
Stability for fixed $t$ & Most stable for large changes & Least stable for small changes & Most stable for small changes, least stable for large changes\tabularnewline
\hline 
Stability for fixed exchange rate derivative & Most stable & Mid-stability & Least stable\tabularnewline
\hline 
\end{tabular}
\par\end{centering}
\caption{Summary of the comparison of the three mixing types.}
\end{table}

\section{\label{sec: Advanced-Construction}Advanced Homotopy Construction}

There are two important methods for creating more advanced and dynamic
behavior in these homotopy AMMs. As an example, it is possible to
think of the parameter $t$ as a function of $x$ and $y$ in order
to create a non-uniform homotopy between $A_{0}(x,y)$ and $A_{1}(x,y)$.
A practical example of this appears for Stableswap in \cite{Ego19}.
Recall that the defining equation there is 
\[
\chi D^{n-1}\sum_{j=1}^{n}x_{j}+\prod_{j=1}^{n}x_{j}=\chi D^{n}+\left(\frac{D}{n}\right)^{n}
\]

\noindent where initial values $X^{i}=(x_{1}^{i},...,x_{n}^{i})$
are chosen such that $x_{j}^{i}=\frac{D}{n}$ for each $j$. Initially
there is the equality $\prod_{j=1}^{n}x_{j}^{i}=\left(\frac{D}{n}\right)^{n}$,
but as the market state falls out of this initial balance the equality
will be less and less exact. The proposed solution to counter this
lack of balance is to dynamically adjust the leverage parameter $\chi$.
For some fixed choice of $A$, the value of of $\chi$ is given by

\[
\chi=A\frac{\prod_{j=1}^{n}x_{j}}{\left(\frac{D}{n}\right)^{n}}
\]

\noindent When this is substituted into the above equation, the curve
is now defined by the following:

\[
An^{n}\sum_{j=1}^{n}x_{j}+D=ADn^{n}+\frac{D^{n+1}}{n^{n}\prod_{j=1}^{n}x_{j}}
\]

To get more specific with this construction, let's take $n=2$ and
turn to the above equations as well as those for the homotopy mixing
curves:

\noindent 
\begin{eqnarray*}
\chi & = & \frac{4Axy}{D^{2}}\\
t & = & \frac{D^{2}}{16Axy+D^{2}}\\
A_{t}^{hom}(x,y) & = & \frac{D(1-t)}{x+y}+\frac{Dt}{2\sqrt{xy}}
\end{eqnarray*}

\noindent The above gives a fixed rule for dynamically updating the
value of $t$ as the state $(x,y)$ of the AMM changes. Substituting
this function for $t$ into $A_{t}^{hom}(x,y)=1$ yields the following
curve:

\[
\frac{16ADxy}{x+y}+\frac{D^{3}}{2\sqrt{xy}}=16Axy+D^{2}
\]

\begin{figure}[t]
\begin{centering}
\begin{tabular}{cc}
\includegraphics[scale=0.5]{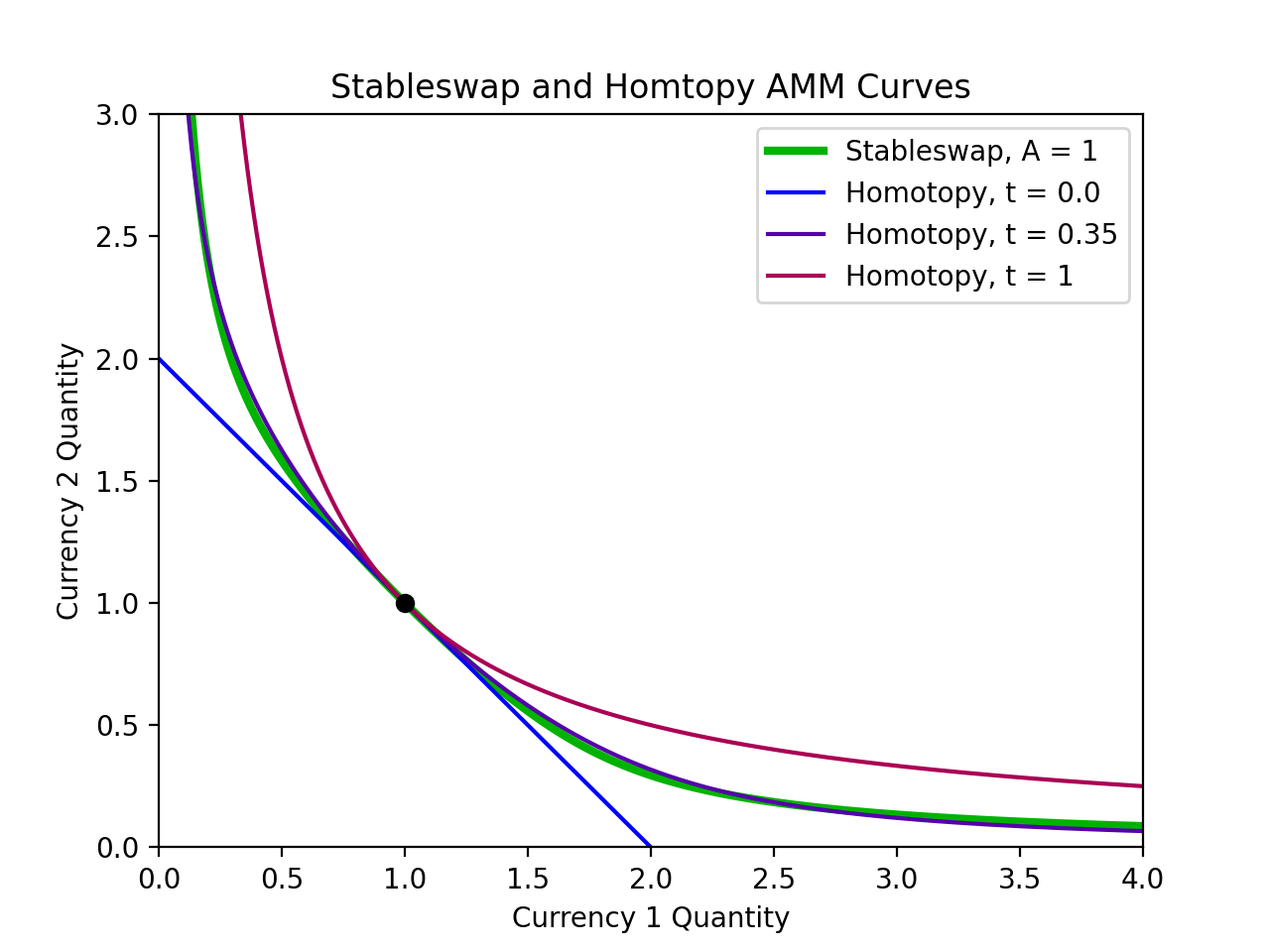} & \includegraphics[scale=0.5]{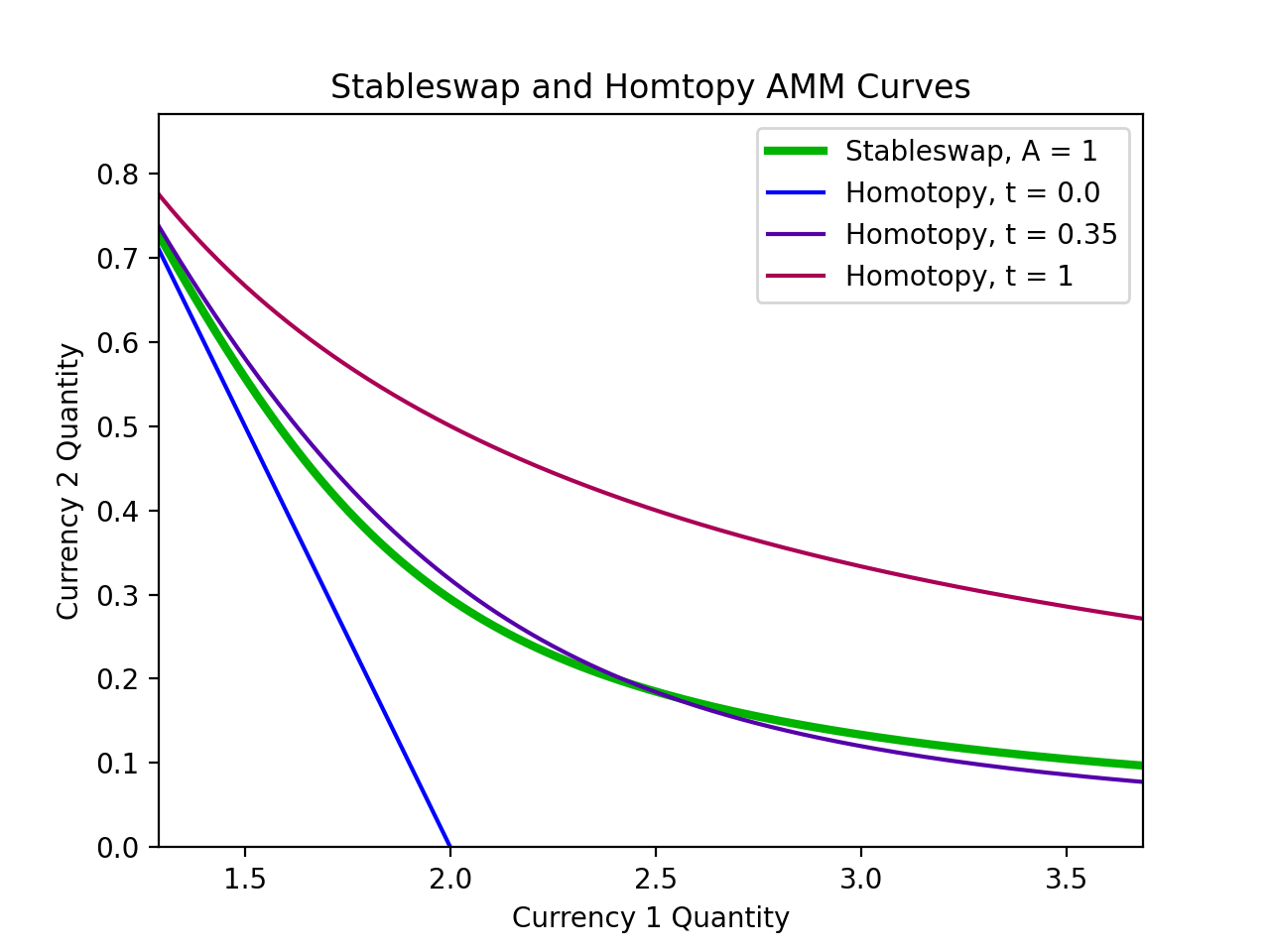}\tabularnewline
\end{tabular}
\par\end{centering}
\caption{\label{fig: Stableswap homotopy curve comparison}Comparison of the
Stableswap non-uniform homotopy curve with a uniform homotopy curve.
The left image is the majority of the curve; the right image is a
zoomed in section that demonstrates how no single value of $t$ captures
the behavior of the Stableswap curve.}

\end{figure}

\noindent As Figure \ref{fig: Stableswap homotopy curve comparison}
demonstrates, this new curve represents a homotopy between the CSMM
and CPMM where the rate of transition from one to the other is non-uniform
over the curve. Note that it is possible to solve for $t$ as a function
of only $s$, but the result is a rather complicated formula and is
not worth presenting here.

\subsection{\label{subsec:Constructing-Specific-Non-Unifor}Constructing Specific
Non-Uniform Homotopy Behavior}

The original definition of the homotopy mixing curve $A_{t}^{hom}(x,y)$
assumed that $t$ was a uniform constant. However, the above Stableswap
construction demonstrates that interesting behavior can appear when
$t$ is non-uniform and is taken to be a function of other variables
(\cite{Ego19}). The purpose of this section is (1) to argue that
taking $t$ to be a function of $s$ is a naturally powerful way for
designing non-uniform homotopy curves, (2) to provide required conditions
on this function $t(s)$ to ensure the resulting curve satisfies the
axioms of an AMM, and (3) to demonstrate examples of interesting curves
using this construction.

In Section \ref{sec: General Construction}, a method for parametrizing
the family of homotopy AMM curves is presented. In particular, a choice
of values of $s$ and $t$ returns a point $(x,y)$ according to the
following equation:

\noindent 
\begin{eqnarray*}
(x,y) & = & \left((ax_{0}+by_{0})(1-t)+\left(\frac{x_{0}}{x}\right)^{\frac{\alpha}{\alpha+\beta}}\left(\frac{y_{0}}{y}\right)^{\frac{\beta}{\alpha+\beta}}t\right)\left(\frac{s}{a},\frac{1-s}{b}\right)\\
 & = & \left(\left(\left(\frac{x_{0}}{x}\right)^{\frac{\alpha}{\alpha+\beta}}\left(\frac{y_{0}}{y}\right)^{\frac{\beta}{\alpha+\beta}}-(ax_{0}+by_{0})\right)t+(ax_{0}+by_{0})\right)\left(\frac{s}{a},\frac{1-s}{b}\right)
\end{eqnarray*}

\[
\implies\lambda(x,y)=\left(\left(\frac{x_{0}}{x}\right)^{\frac{\alpha}{\alpha+\beta}}\left(\frac{y_{0}}{y}\right)^{\frac{\beta}{\alpha+\beta}}-(ax_{0}+by_{0})\right)t+(ax_{0}+by_{0})
\]

\noindent This idea is further developed using the fact that $s=\frac{ax}{ax+by}$,
meaning that an entire curve is simply given by a single value of
$t$. The above introduction discusses how one can think of the Stableswap
curve in this context by taking $t$ to be a function of $x$ and
$y$. However, this section reduces the system to a single parameter
by taking $t$ to be a function of $s$ directly. Of course, because
$s$ can be determined by $x$ and $y$ the result is practically
no different. The advantage of this new perspective it that it helps
in designing more predictable behavior.

\noindent 
\begin{eqnarray*}
A^{hom}(x,y) & = & \left(\frac{ax_{0}+by_{0}}{ax+by}\right)\left(1-t\left(\frac{ax}{ax+by}\right)\right)+\left(\frac{x_{0}}{x}\right)^{\frac{\alpha}{\alpha+\beta}}\left(\frac{y_{0}}{y}\right)^{\frac{\beta}{\alpha+\beta}}t\left(\frac{ax}{ax+by}\right)\\
 & = & \left(\left(\frac{x_{0}}{x}\right)^{\frac{\alpha}{\alpha+\beta}}\left(\frac{y_{0}}{y}\right)^{\frac{\beta}{\alpha+\beta}}-\frac{ax_{0}+by_{0}}{ax+by}\right)t\left(\frac{ax}{ax+by}\right)+\frac{ax_{0}+by_{0}}{ax+by}
\end{eqnarray*}

In the development of this paper, it became apparent that even simple
equations for $t(s)$ can sometimes fail to provide proper AMMs; the
repeated issue is that the curve can fail to be convex but it was
unclear why this was happening. The goal here is to provide a simple
condition on $t(s)$ to ensure that the resulting AMM is convex. The
second derivative of this curve must be computed to find the necessary
condition because a twice differentiable function is convex if and
only if its second derivative is non-negative.

To find this second derivative, note that the above equation for $A^{hom}(x,y)$
gives an implicit definition of the curve. In general it is not possible
to solve for $y$ as a function of $x$, but it is still possible
to implicitly compute the first and second derivatives of this homotopy
curve. For these computations think of $x$ and $y$ as functions
of $s$ and use the following shorthand for simplicity:

\noindent 
\begin{eqnarray*}
C=ax_{0}+by_{0} &  & P(s)=\left(\frac{ax_{0}}{s}\right)^{\frac{\alpha}{\alpha+\beta}}\left(\frac{by_{0}}{1-s}\right)^{\frac{\beta}{\alpha+\beta}}
\end{eqnarray*}

\noindent Next, denoting $\lambda(s)=\lambda^{hom}(s)$ and differentiating
yields the following:

\noindent 
\begin{eqnarray*}
\lambda(s) & = & (P(s)-C)t(s)+C\\
\\
\lambda'(s) & = & (P(s)-C)t'(s)+P'(s)t(s)\\
 & = & (P(s)-C)t'(s)-\frac{\alpha(1-s)-\beta s}{s(1-s)(\alpha+\beta)}P(s)t(s)\\
\\
\lambda''(s) & = & (P(s)-C)t''(s)+2P'(s)t'(s)+P''(s)t(s)\\
 & = & (P(s)-C)t''(s)-2\frac{\alpha(1-s)-\beta s}{s(1-s)(\alpha+\beta)}P(s)t'(s)\\
 &  & +\frac{2\alpha^{2}(1-s)^{2}+\alpha\beta(1-2s)^{2}+2\beta^{2}s^{2}}{s^{2}(1-s)^{2}(\alpha+\beta)^{2}}P(s)t(s)
\end{eqnarray*}

\noindent The functions $x=x(s)$ and $y=y(s)$ and their derivatives
can now be written in terms of $\lambda$ and its derivatives:

\noindent 
\begin{eqnarray*}
x(s)=\frac{s}{a}\lambda(s) &  & y(s)=\frac{1-s}{b}\lambda(s)\\
x'(s)=\frac{1}{a}\lambda(s)+\frac{s}{a}\lambda'(s) &  & y'(s)=-\frac{1}{b}\lambda(s)+\frac{1-s}{b}\lambda'(s)\\
x''(s)=\frac{2}{a}\lambda'(s)+\frac{s}{a}\lambda''(s) &  & y''(s)=-\frac{2}{b}\lambda'(s)+\frac{1-s}{b}\lambda''(s)
\end{eqnarray*}

\noindent One can show that the following equations hold as definitions
for implicit derivatives in this case:

\noindent 
\begin{eqnarray*}
\frac{dy}{dx}(s) & = & \frac{y'(s)}{x'(s)}\\
\frac{d^{2}y}{dx^{2}}(s) & = & \frac{x'(s)y''(s)-x''(s)y'(s)}{x'(s)^{3}}
\end{eqnarray*}

\noindent The equations for $x$ and $y$ in terms of $\lambda$ can
be substituted in here and result can be simplified:

\noindent 
\begin{eqnarray*}
\frac{dy}{dx}(s) & = & \left(\frac{\lambda'(s)}{\lambda(s)+s\lambda'(s)}-1\right)\cdot\frac{a}{b}\\
\frac{d^{2}y}{dx^{2}}(s) & = & \frac{\lambda(s)\lambda''(s)-2\lambda'(s)^{2}}{(\lambda(s)+s\lambda'(s))^{3}}\cdot\frac{a^{2}}{b}
\end{eqnarray*}

\noindent Note that $b$ and $x'(s)$ should always be positive by
the construction of the homotopy curve as an AMM. Thus, under these
assumptions the above can be distilled into a simple convexity condition
in terms of $\lambda$:

\begin{equation}
A^{hom}(x,y)=1\text{\thinspace\thinspace\thinspace is convex if and only if\thinspace\thinspace\thinspace\ensuremath{\lambda(s)\lambda''(s)\geq2\lambda'(s)^{2}\text{\thinspace\thinspace\thinspace for all \ensuremath{s\in(0,1)}}}}
\end{equation}

\begin{figure}[t]
\begin{centering}
\begin{tabular}{cc}
\includegraphics[scale=0.5]{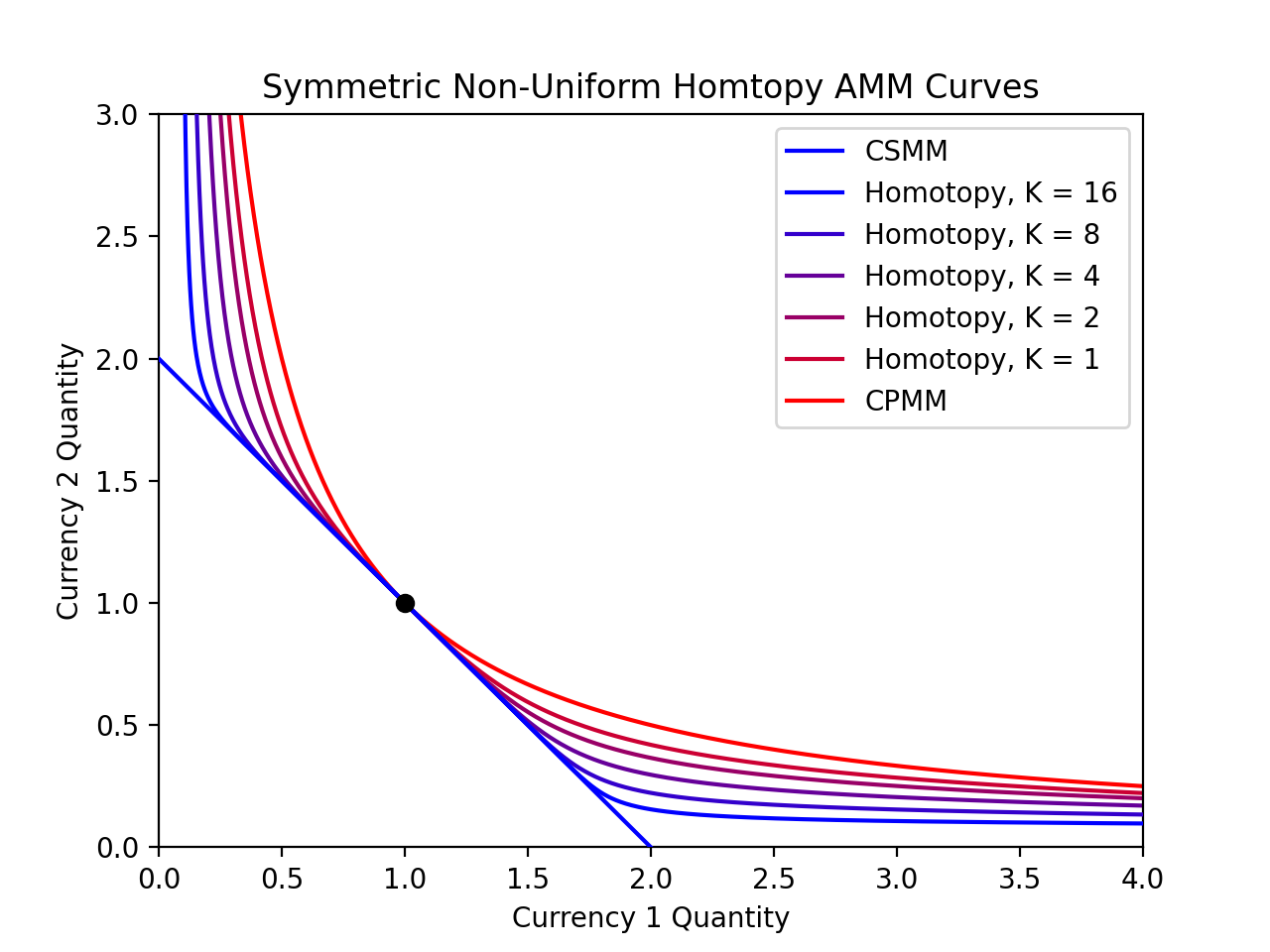} & \includegraphics[scale=0.5]{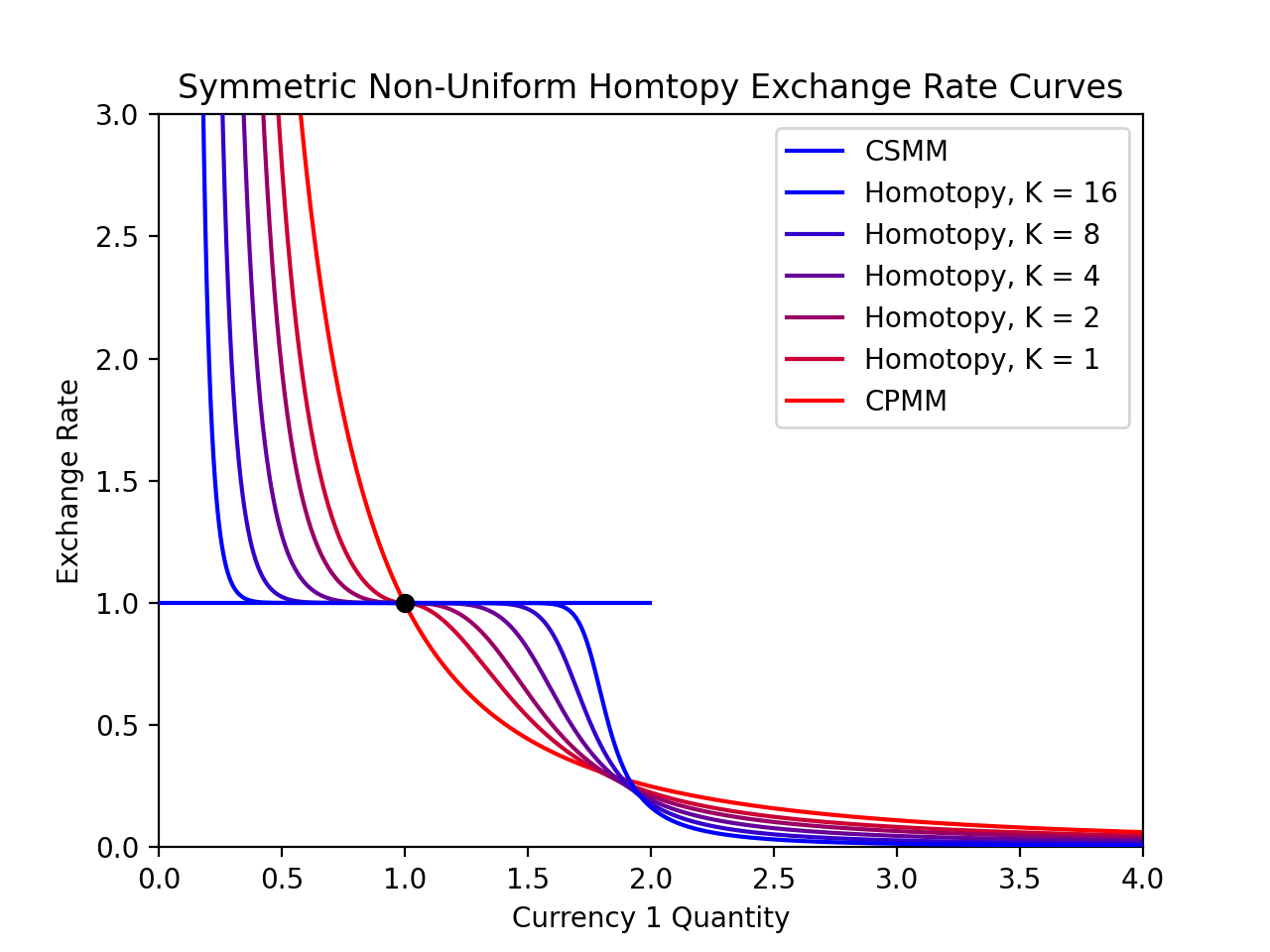}\tabularnewline
\end{tabular}
\par\end{centering}
\caption{\label{fig:Examples-of-symmetric}Examples of symmetric non-uniform
homotopy curves and their corresponding exchange rates. The parameter
$K$ is thought of as a stability parameter where a larger value of
$K$ forces the curve to more closely resemble a CSMM. When comparing
to the homotopy curves of Figure \ref{fig: curve time comparisons},
it is clear that the advantage of this non-uniformity is a much more
stable exchange rate given changes in the quantities of currency.}
\end{figure}

\begin{figure}[t]
\begin{centering}
\begin{tabular}{cc}
\includegraphics[scale=0.5]{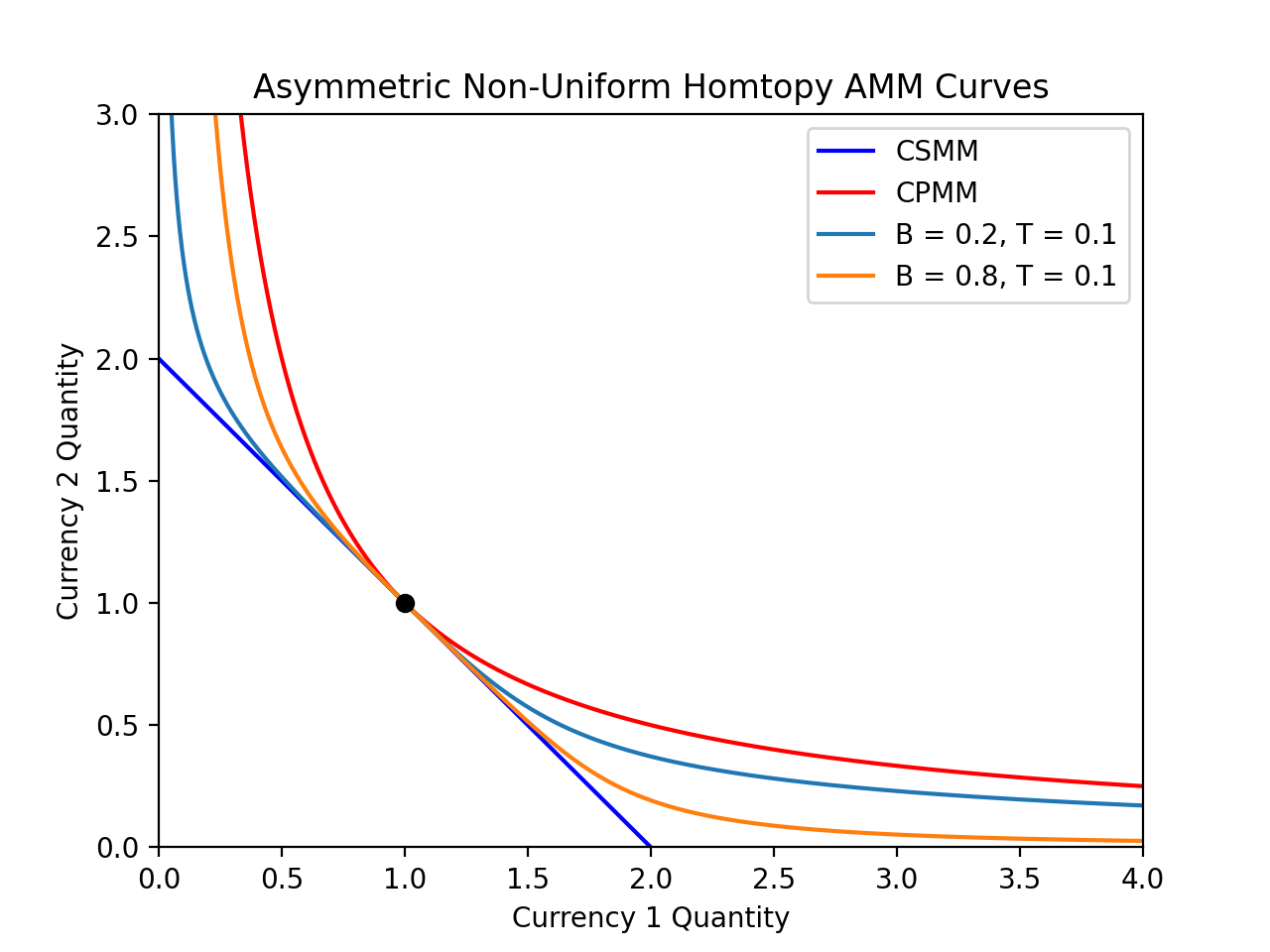} & \includegraphics[scale=0.5]{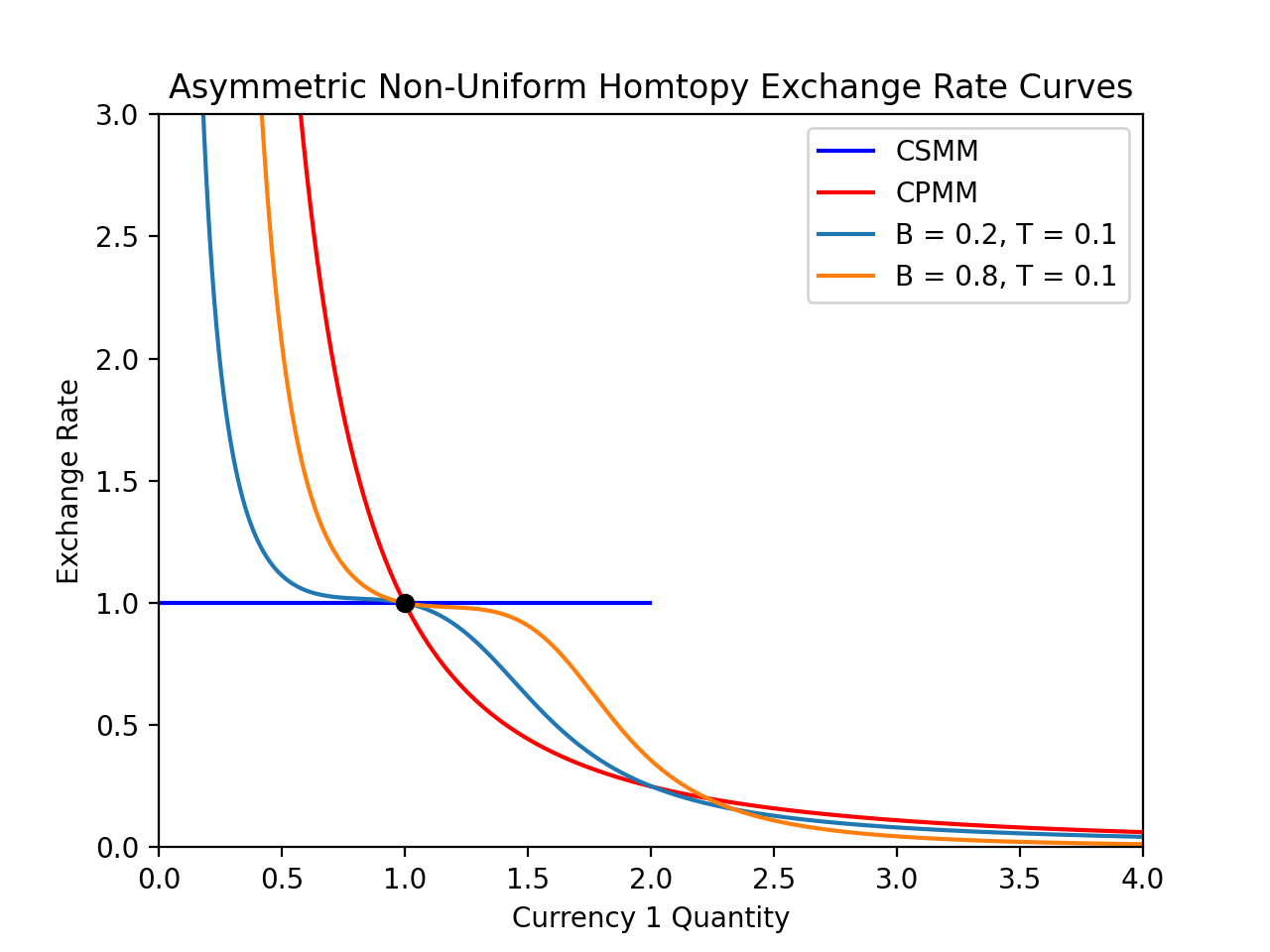}\tabularnewline
\includegraphics[scale=0.5]{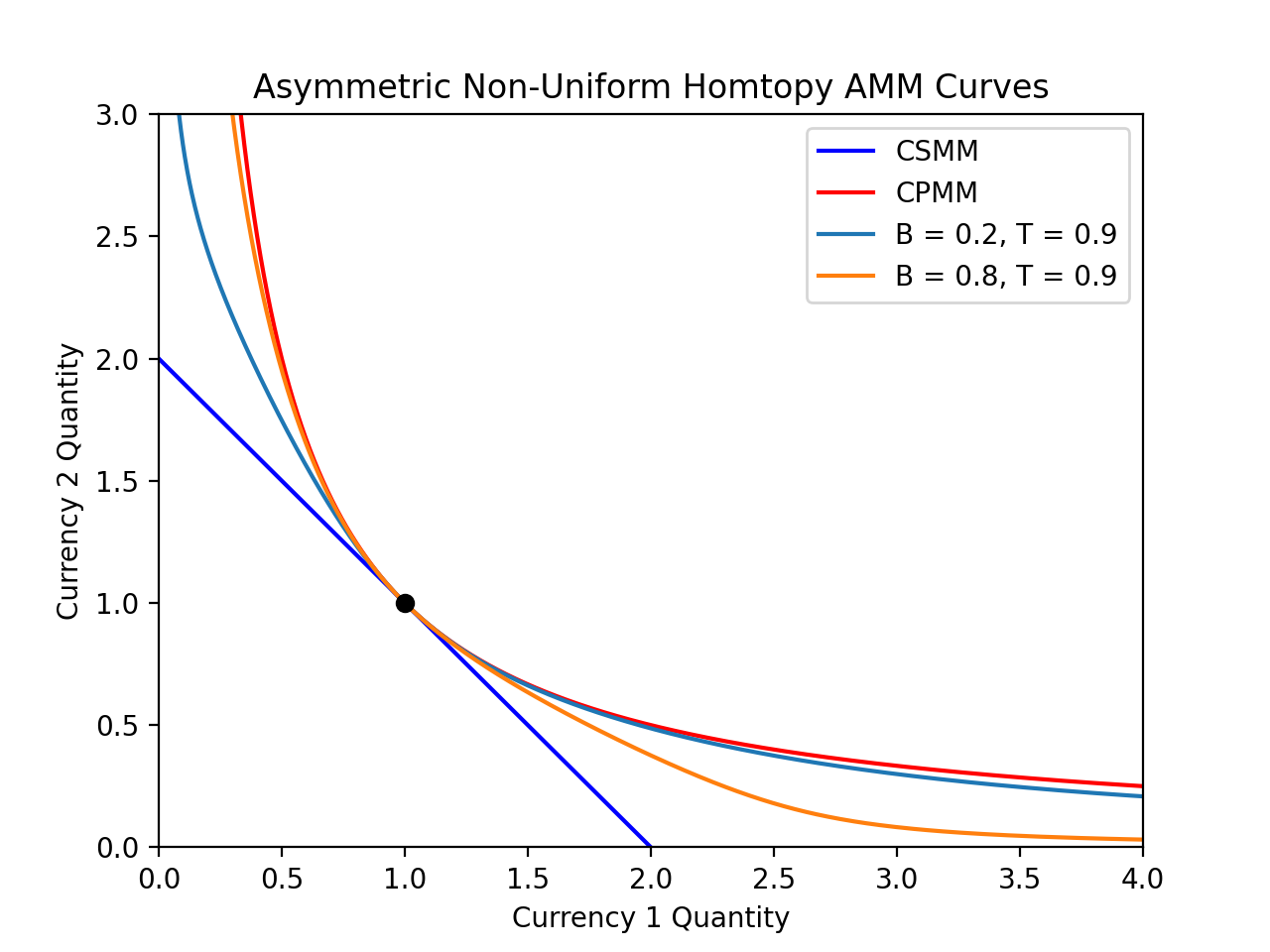} & \includegraphics[scale=0.5]{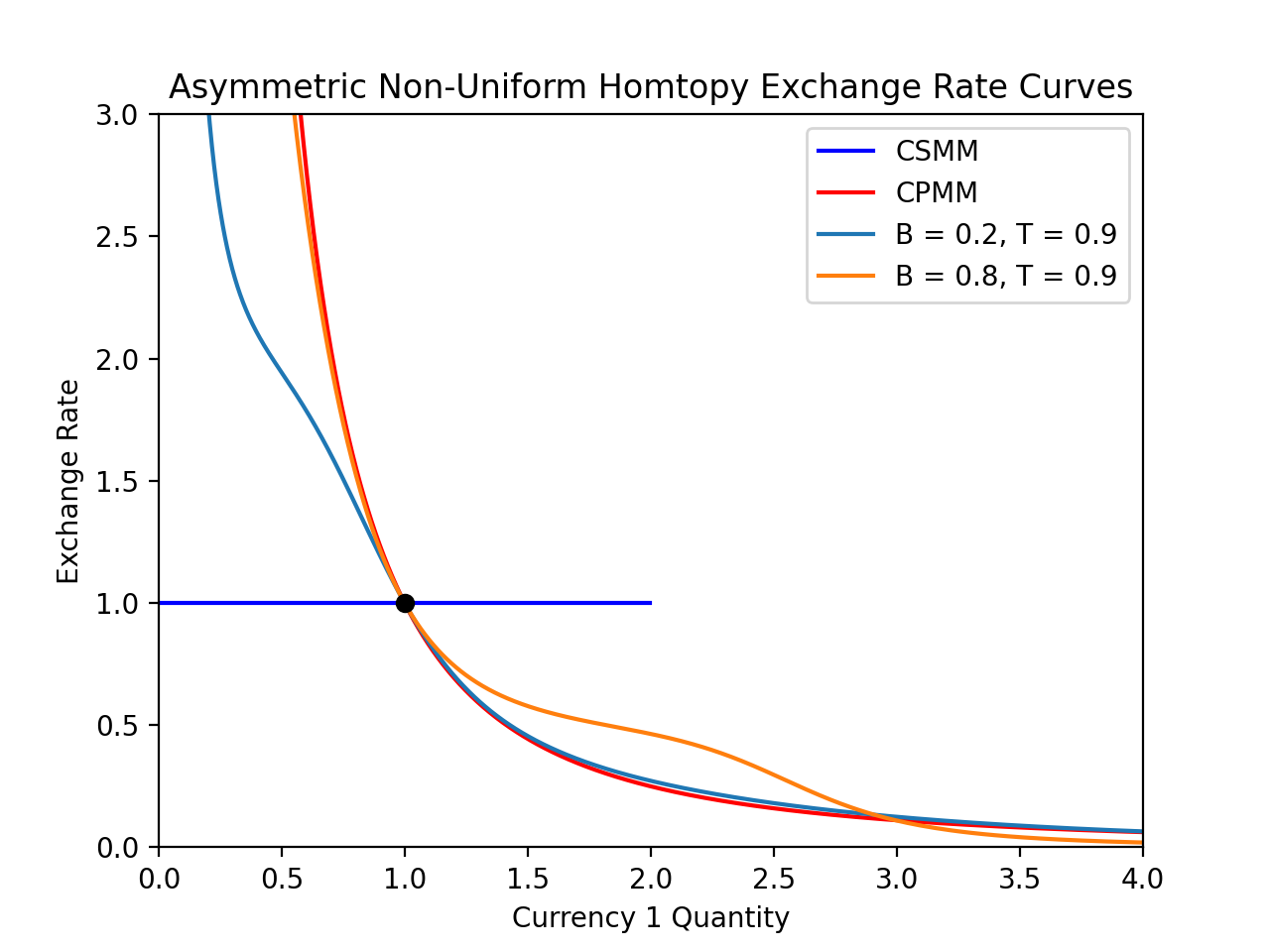}\tabularnewline
\end{tabular}
\par\end{centering}
\caption{\label{fig:Examples-of-asymmetric}Examples of asymmetric non-uniform
homotopy curves and their corresponding exchange rate curves. (Left)
The AMM curves demonstrate that a high value of the bias parameter
$B$ creates a curve where the behavior more closely resembles a CSMM
when the quantity of Currency 1 is higher than the initial quantity.
In addition, the parameter $T$ controls the overall transition from
CSMM to CPMM as the value of $t$ does in the uniform homotopy case.
(Right) The exchange rate curves demonstrate that the bias parameter
controls the exchange rate near the initial point. In particular,
the exchange rates to both the left and right of the initial point
are lower for the $B=0.2$ curve than they are for the $B=0.8$ curve.}
\end{figure}

The following is a summary of some basic examples of non-uniform homotopy
curves. These examples present a novel way to create specifically
designed behavior in the stability and overall shape of the curve.
For these curves, take $s_{0}=\frac{ax_{0}}{ax_{0}+by_{0}}$ to be
the value of $s$ corresponding to the initial point $(x_{0},y_{0})$.
\begin{enumerate}
\item Denote $M=max(s_{0},1-s_{0})$. Take $K$ to be a ``stability''
parameter, where $K\geq0$ and larger values of $K$ give curves with
higher price stability. Then the following family of functions are
always valid in that the resulting $A^{hom}$ curves are convex:

\[
t(s)=\left|\frac{s-s_{0}}{M}\right|^{K}
\]

\item Let $B$ be a ``bias'' parameter where $t(0)=B$ and $t(1)=1-B$.
Let $T$ be a parameter controlling $t$ at $s=s_{0}$, i.e. $t(s_{0})=T$.
A parabolic extension of this yields the following:

\[
t(s)=\frac{(1-2B)s_{0}+(B-T)}{s_{0}(1-s_{0})}s^{2}-\frac{(1-2B)s_{0}^{2}+(B-T)}{s_{0}(1-s_{0})}s+B
\]

The values of $B$ and $T$ must be chosen so that $0\leq t(s)\leq1$
and so that the test $\lambda(s)\lambda''(s)\geq2\lambda'(s)^{2}$
will hold for all $s\in(0,1)$; the equations for this are complicated
to write down, but these values can be easily verified with simple
code.
\end{enumerate}
\noindent These curves are visualized in Figures \ref{fig:Examples-of-symmetric}
and \ref{fig:Examples-of-asymmetric} respectively.

\subsection{\label{subsec:Dynamic-Adjustments-to}Dynamic Adjustments to Match
External Prices}

One key point of this construction is that the exchange rate is designed
to be stabilized about the exchange rate of the initial point. There
are other constructions that do this as well, particularly in the
area of AMMs used for stablecoins. For example, the Curve v1 system
from the Stableswap white paper \cite{Ego19} was designed to always
be stabilized to any equal amount of each currency; this is seen by
the fact that each value is initialized to $x_{j}^{i}=\frac{D}{n}$.
The follow up paper \cite{Ego21} for Curve v2 allows for stabilizing
around arbitrary states.

One potential function to add into this homotopy model is the ability
to dynamically adjust the stabilized exchange rate of the model. Recall
that the initial state of the model is given by $(x_{0},y_{0})$ and
that the initial internal exchange rate of the price of Currency 1
in terms of Currency 2 is given by $\frac{a}{b}$. If the the current
state of the AMM given by $A^{hom}(x,y)=1$ is $(x_{1},y_{1})$ then
the current internal exchange rate of the system is given by the following:

\[
\text{current internal exchange rate\thinspace\thinspace\thinspace}=\frac{\frac{\partial A^{hom}}{\partial x}(x_{1},y_{1})}{\frac{\partial A^{hom}}{\partial y}(x_{1},y_{1})}
\]

\noindent In principal, the internal exchange rate of the system is
driven by the external exchange rate; arbitrageurs will tend to make
exchanges biased toward shifting the two exchange rates together.
If the current internal exchange rate gets too far away from the initial
exchange rate then the system could become unstable. One way to try
to relieve this pressure is to update the curve itself; one could
(1) update the initial point to be $(x_{1},y_{1})$, (2) choose new
$a$ and $b$ so that $\frac{a}{b}$ is equal to the external exchange
rate, and (3) update the values of $\alpha$ and $\beta$ accordingly.

As a note before proceeding, the notion of slippage is used precisely
here and should be defined. Let $p_{1}$ denote the current internal
exchange rate or ``estimated spot price''. Suppose a transaction
in the AMM moves the state from $(x_{1},y_{1})$ to $(x_{2},y_{2})$.
The ``actual spot price'' would then be given by $p_{2}=-\frac{y_{2}-y_{1}}{x_{2}-x_{1}}$.
Slippage here is defined to be

\begin{equation}
\text{slippage}=\left|\frac{\text{estimated spot price}-\text{actual spot price}}{\text{estimated spot price}}\right|=\left|\frac{p_{1}-p_{2}}{p_{1}}\right|
\end{equation}

\noindent In other words, slippage is taken to be the magnitude of
the percent difference between the estimated and actual spot prices
of the transaction. This definition is used because the notion of
stability for a homotopy curve is the ability of the curve to maintain
the initial exchange rate given small changes in currency quantity
in the AMM; a very stable curve should have very small differences
in spot prices when in the stable region of the curve.

Simulations show that a curve with high price stability will reduce
slippage when near the stable point but also make it more difficult
for the AMM to match the external exchange rate. The goal of these
simulations was to have a very rough approximation of arbitrageurs
attempting to match external exchange rates in a non-uniform homotopy
AMM. Note that the exact parameters for this simulation are fairly
simplistic and arbitrary; however, changing these parameters did not
yield significantly different takeaways. These simulations were run
with the following conditions:
\begin{itemize}
\item each simulation ran for $500$ time steps
\item the initial state was always set to be $(3000,1000)$
\item a randomly generated external exchange rate curve is reused every
simulation where
\begin{itemize}
\item the initial exchange rate was set to $0.5$
\item the exchange rate could increase or decrease by up to $20\%$ every
$80$ time steps
\end{itemize}
\item $t(s)=\left|\frac{s-s_{0}}{M}\right|^{K}$ was used where $K$ was
equal to $8$ times a stability parameter between $0$ and $1$
\item users could extract up to $2\%$ of the total amount of that currency
in the AMM with each transaction
\item transactions favored moving the internal exchange rate toward the
external exchange rate with a likelihood of $90\%$
\item the underlying curve parameters where never updated over the course
of the simulation
\end{itemize}
\noindent A future goal for this work is to create more sophisticated
simulations with a more in-depth analysis of non-uniform homotopy
curve behavior.

\begin{figure}[t]
\begin{centering}
\begin{tabular}{ccc}
\includegraphics[scale=0.3]{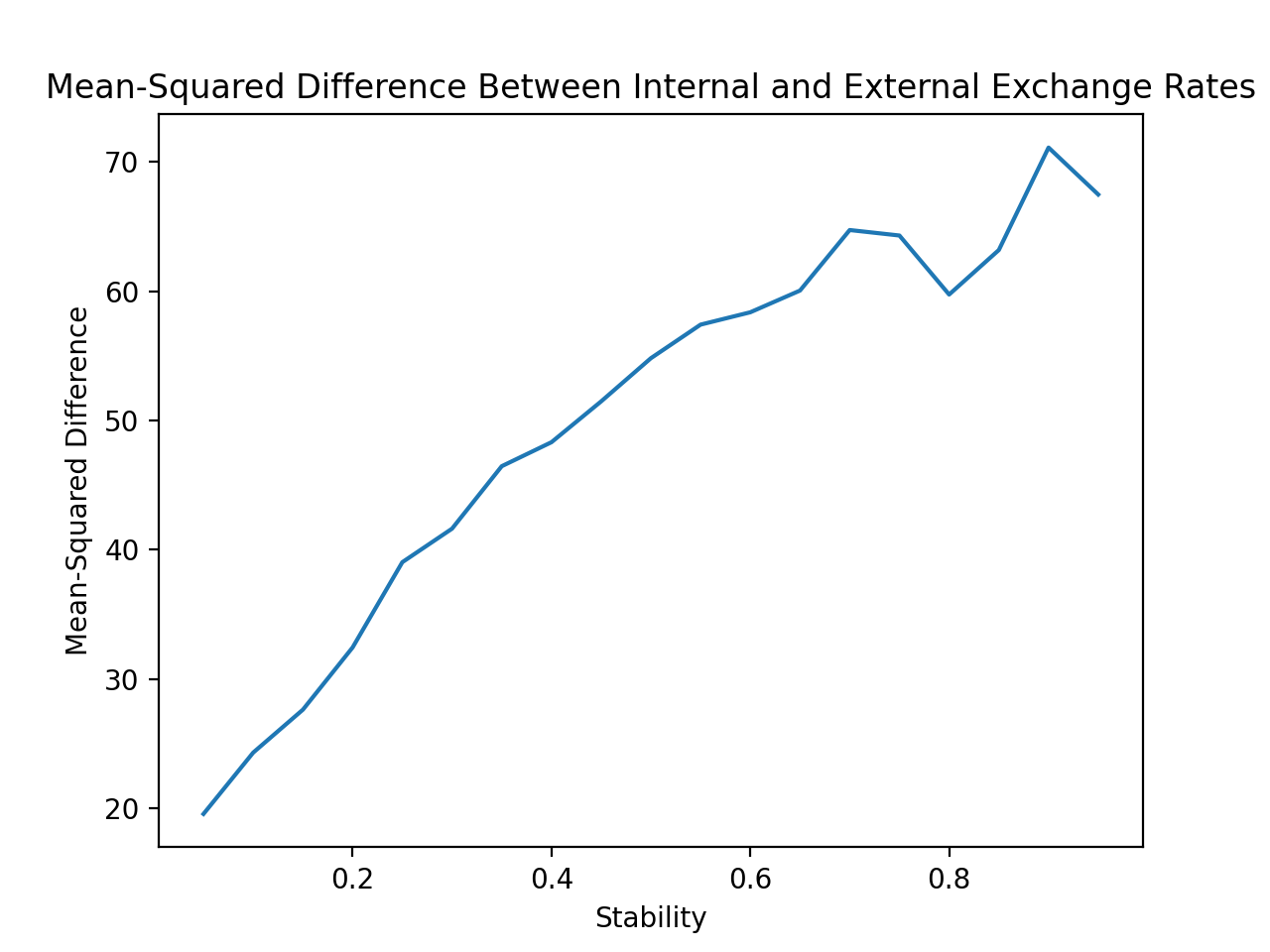} & \includegraphics[scale=0.3]{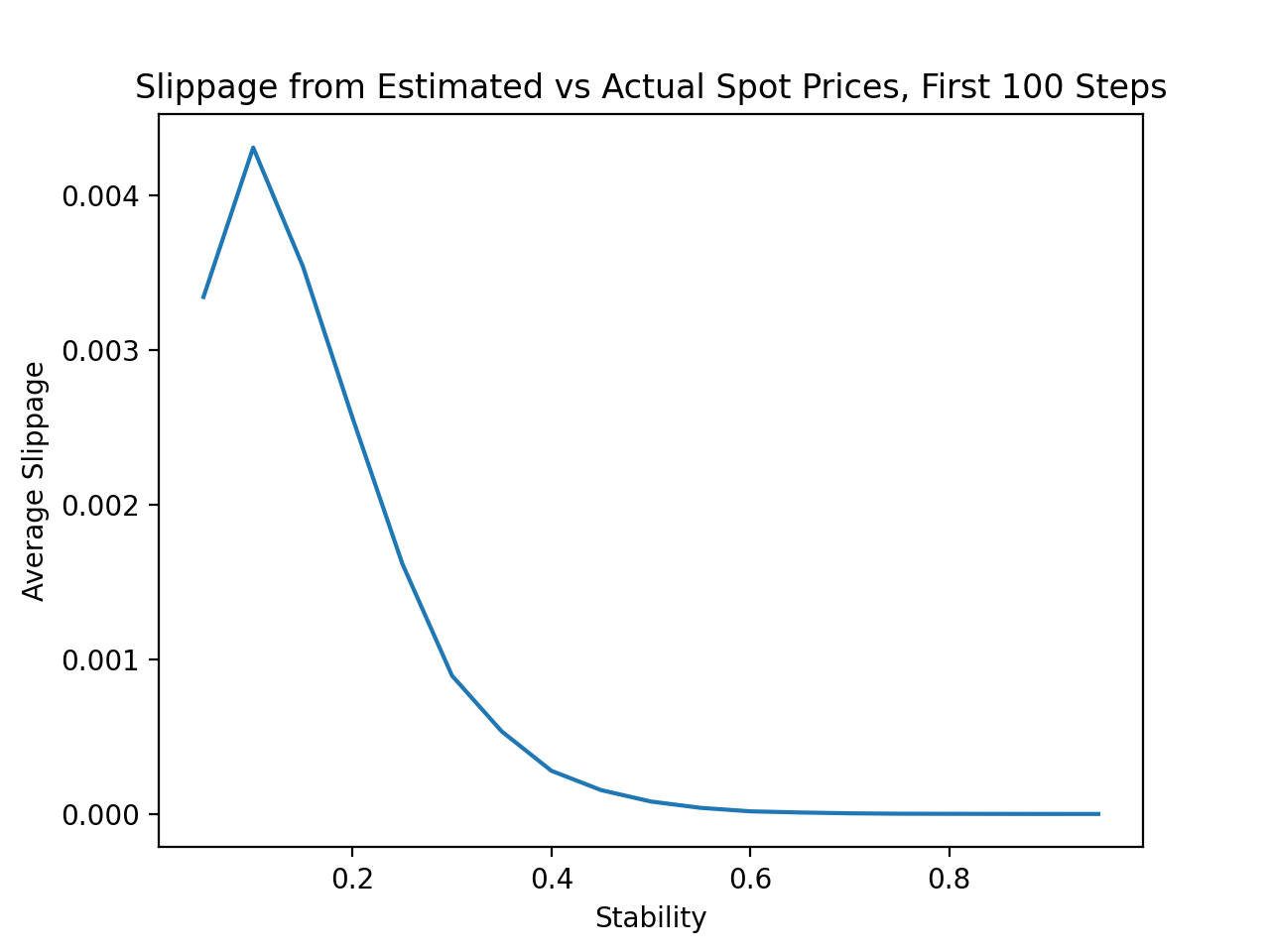} & \includegraphics[scale=0.3]{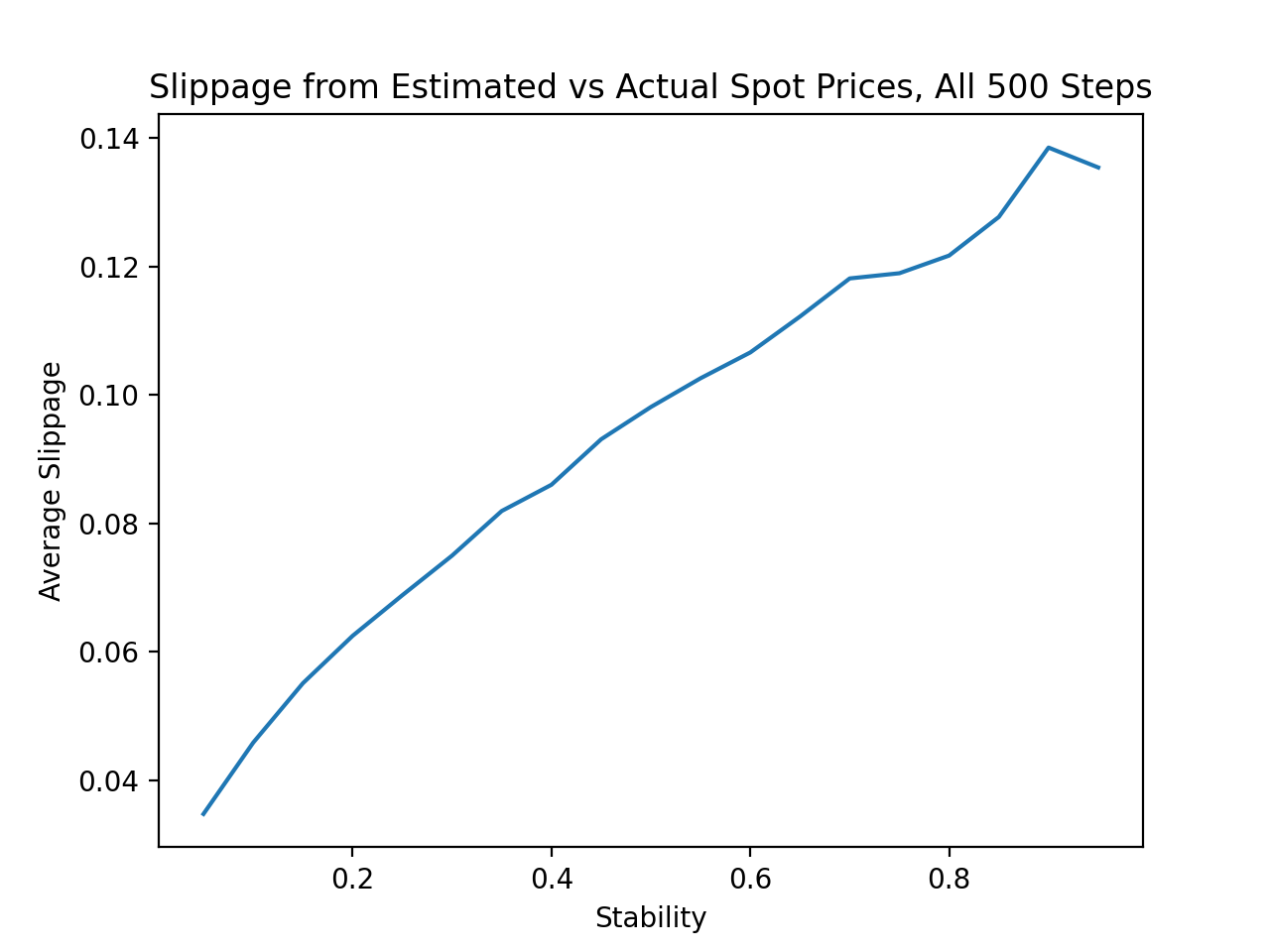}\tabularnewline
\end{tabular}
\par\end{centering}
\caption{The stabilities used were all multiples of $0.05$ between $0.05$
and $0.95$ inclusive. Each stability was run $100$ times and the
results were averaged to create this figure. (Left) Mean-squared difference
between the internal and external rates as a function of the stability.
This demonstrates that the arbitrageurs have a harder time matching
external exchange rates when the stability of the AMM is high. (Center)
The quantities and exchange rate remain close to the initial values
in the first $100$ time steps. Higher price stability in this region
of the curve means that price slippage is lower. (Right) The homotopy
curve model can support any exchange rate, but stability is lost as
the exchange rate diverges from the initial one. By time step $500$
the external exchange rate has changed too much and the stability
is lost, and this loss is more prevalent the higher this original
stability was; this is an argument for why dynamically adjusting the
curve is important.}
\end{figure}

\section{Portfolio Value Functions and Stability}

As introduced in \cite{AEC21} and discussed in \cite{TPL22}, there
is an elegant mathematical relationship between AMMs and their corresponding
portfolio value functions. For completeness, an adapted version of
that work is presented here. Suppose $P=(p_{1},...,p_{n})$ is a price
vector and the AMM is defined by $A(x_{1},...,x_{n})=k$ using the
function $A:\mathbb{R}_{>0}^{n}\rightarrow\mathbb{R}$. The value
of the portfolio at the price vector $P$ is defined to be

\begin{equation}
V(P)=inf\{P\cdot X:A(X)=k\}
\end{equation}

\noindent The idea of this definition is that arbitrageurs will extract
value from the market by making transactions that adjust the currency
balances; this process continues until the value remaining in the
market has been minimized and the arbitrageurs can't extract any more.

There are a couple observations to make regarding the properties of
this function. First, the use of Lagrange multipliers shows that if
the function $A$ is differentiable then $P$ will be parallel to
$\nabla A(X)$; i.e. one must find the point on the AMM where the
price vector is normal to the surface at that point. This provides
a nice geometric description of the solution to this minimization
problem. Second, as pointed out in \cite{TPL22}, one may think of
the portfolio value function as being related to a Legendre transformation.
One definition of this transformation is as follows. Suppose $f:(a,b)\rightarrow\mathbb{R}$
is a differentiable function; the Legendre transformation of $f$
(if it exists) is denoted by $f^{*}$ and satisfies the property that
$(f^{*})'$ is the inverse of $f'$. Note that convexity of $f$ is
enough to guarantee that such a transformation exists. To see how
portfolio value functions and Legendre transformations are related
by an example:
\begin{itemize}
\item Suppose $f:(a,b)\rightarrow\mathbb{R}$ satisfies the property $A(x,f(x))=k$
\item The value at price vector $(p,q)$ is given by

\begin{eqnarray*}
V(p,q) & = & inf\{px+qy:A(x,y)=k\}\\
 & = & inf\{px+qf(x):x\in(a,b)\}
\end{eqnarray*}

\item In order to minimize the function $g(x)=px+qf(x)$:

\[
g'(x)=p+qf'(x)=0\implies f'(x)=-\frac{p}{q}
\]

\item By assumption $f'$ is invertible and therefore

\[
x=(f')^{-1}\left(-\frac{p}{q}\right)=(f^{*})'\left(-\frac{p}{q}\right)
\]

\item Denoting $\mathcal{L}=f^{*}$ for simplicity, $V(p,q)$ can be written
using the Legendre transformation:

\begin{equation}
V(p,q)=p\mathcal{L}'\left(-\frac{p}{q}\right)+qf\left(\mathcal{L}'\left(-\frac{p}{q}\right)\right)
\end{equation}

\end{itemize}
These two notions are closely related in a broader context for more
general AMMs but that material will not be presented here.

One relevant simplification is made to the portfolio value function
in \cite{AEC21} and \cite{TPL22}. It is clear that $V(P)$ is $1$-homogeneous
because if $\lambda>0$ then

\noindent 
\begin{eqnarray*}
V(\lambda P) & = & inf\{\lambda P\cdot X:A(X)=k\}\\
 & = & \lambda\cdot inf\{P\cdot X:A(X)=k\}\\
 & = & \lambda V(P)
\end{eqnarray*}

\noindent Thus, it makes sense to rewrite the function in a way that
extracts only the relevant information. The reduced portfolio value
function $U:\mathbb{R}_{>0}^{n-1}\rightarrow\mathbb{R}$ is defined
to satisfy the following property:

\begin{equation}
V(p_{1},...,p_{n})=p_{n}U\left(\frac{p_{1}}{p_{n}},...,\frac{p_{n-1}}{p_{n}}\right)
\end{equation}

\begin{figure}[t]
\begin{centering}
\begin{tabular}{ccc}
\includegraphics[scale=0.3]{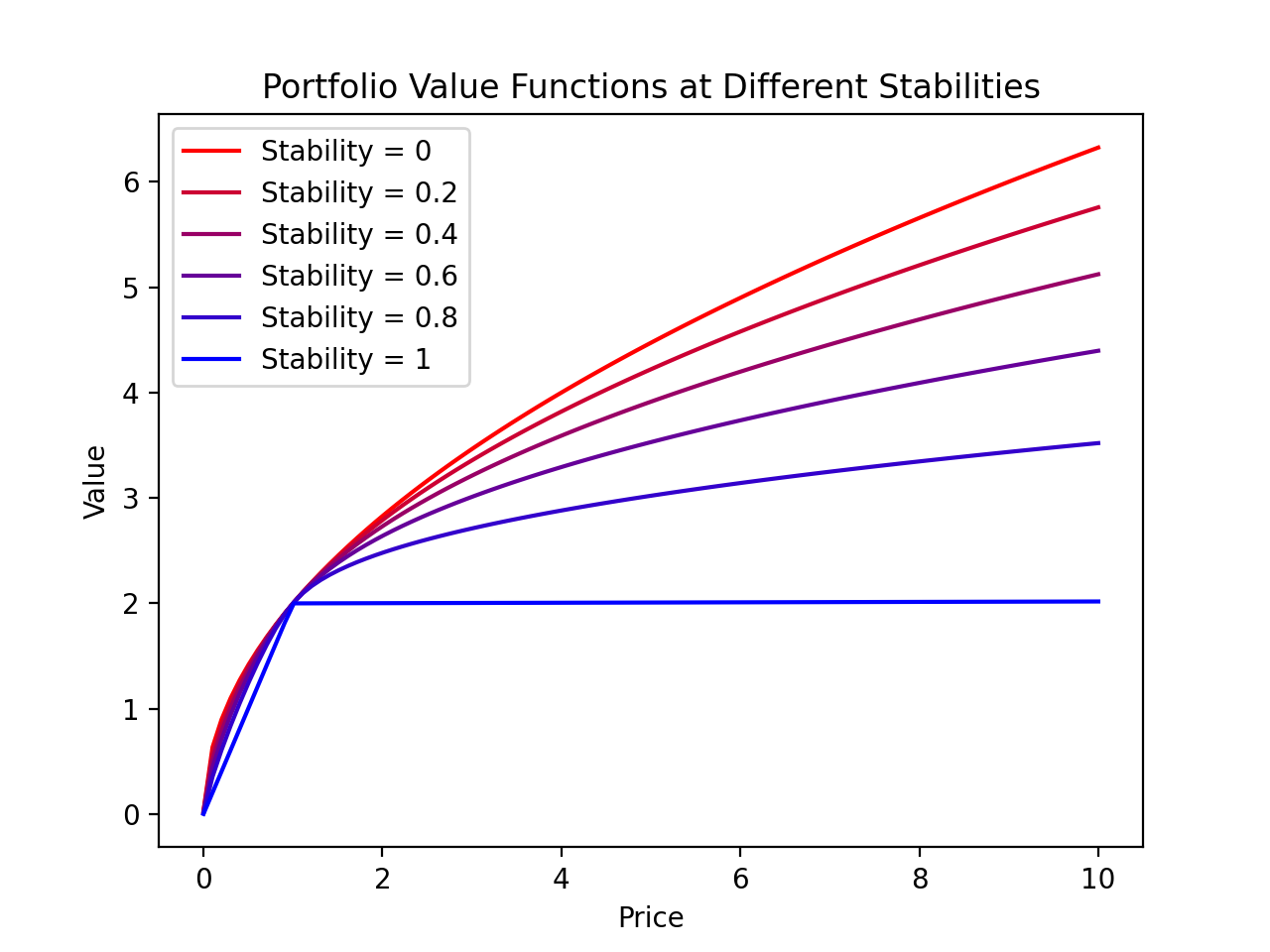} & \includegraphics[scale=0.3]{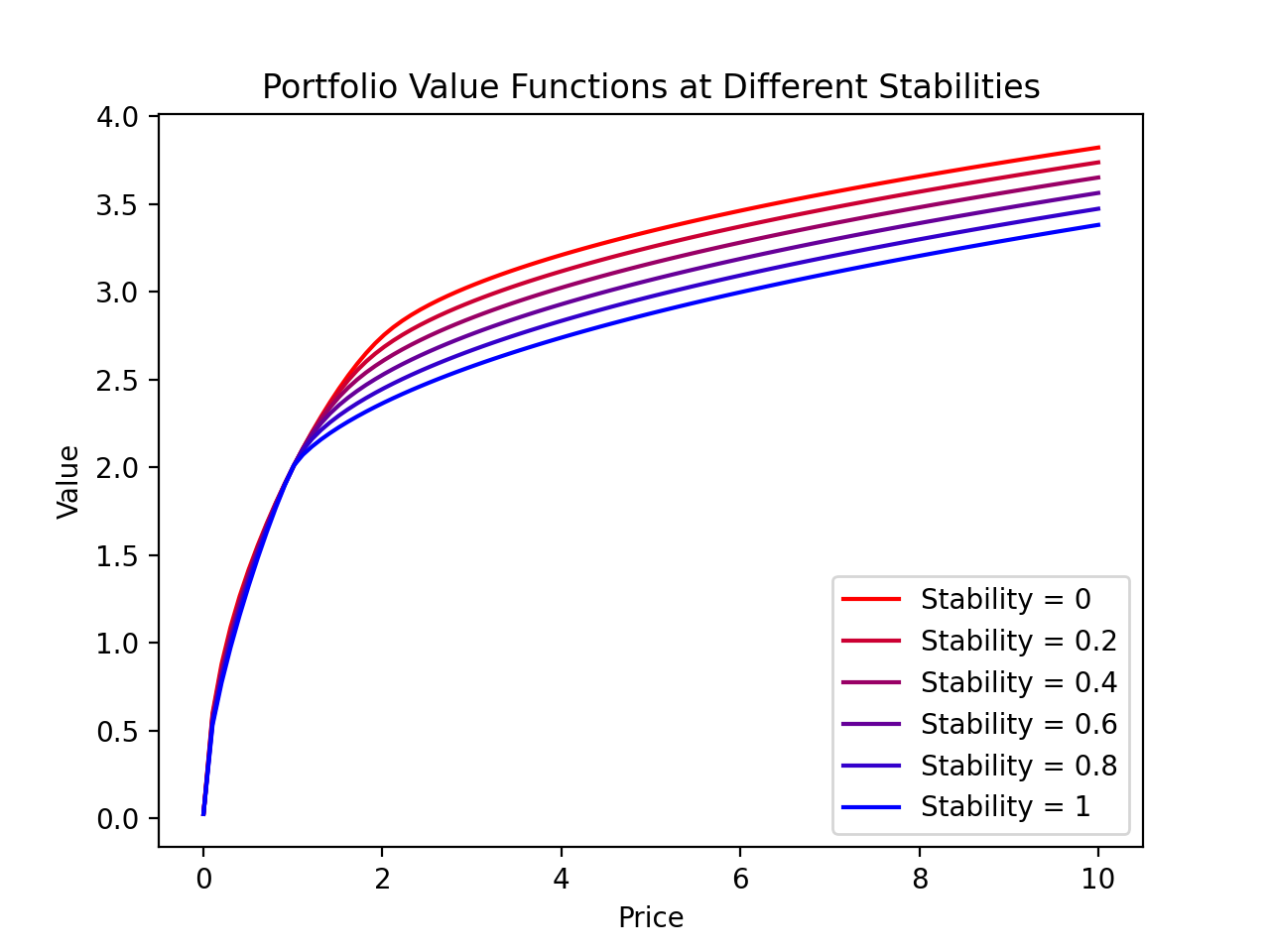} & \includegraphics[scale=0.3]{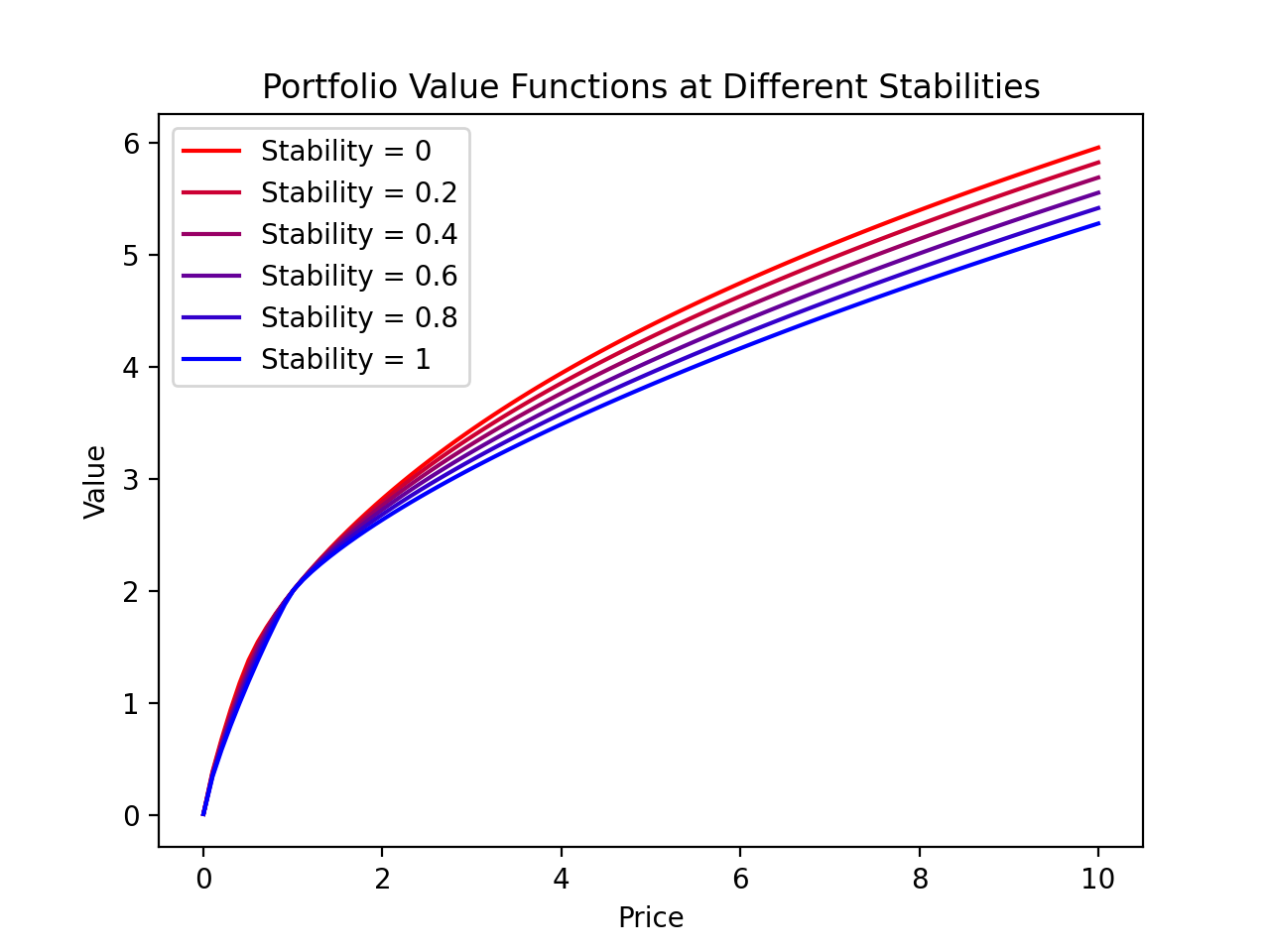}\tabularnewline
\end{tabular}
\par\end{centering}
\caption{\label{fig: portfolio value functions}(Left) Reduced portfolio value
functions of uniform homotopy curves $A_{t}^{hom}(x,y)=1$ for different
stability values. Note that stability here is equal to $1-t$; thus,
stabilities of $0$ and $1$ correspond to CPMM and CSMM respectively.
(Middle and Right) Portfolio value functions of asymmetric non-uniform
homotopy curves $A^{hom}(x,y)=1$ where $t(s)=\frac{(1-2B)s_{0}+(B-T)}{s_{0}(1-s_{0})}s^{2}-\frac{(1-2B)s_{0}^{2}+(B-T)}{s_{0}(1-s_{0})}s+B$.
The middle and right figures have $B=0.2$ and $0.8$ respectively.
The value of $T$ is given by $0.9*(1-\text{stability})+0.1*\text{stability}$.
These curves demonstrate that (1) an increase in stability causes
a decrease in value and (2) an increase in bias causes an increase
in value.}

\end{figure}

\noindent In other words, $V$ is defined in terms of prices $p_{i}$
while $U$ is defined in terms of exchange rates $r_{i}$ (where the
$n^{th}$ currency is thought of as the numéraire) (\cite{AEC21}).
It is clear that $U(r_{1},...,r_{n-1})=V(r_{1},...,r_{n-1},1)$ and
therefore this offers the convenience of being easily graphed in the
$2$-dimensional case (see Figure \ref{fig: portfolio value functions}).

\begin{figure}[t]

\begin{centering}
\includegraphics[scale=0.75]{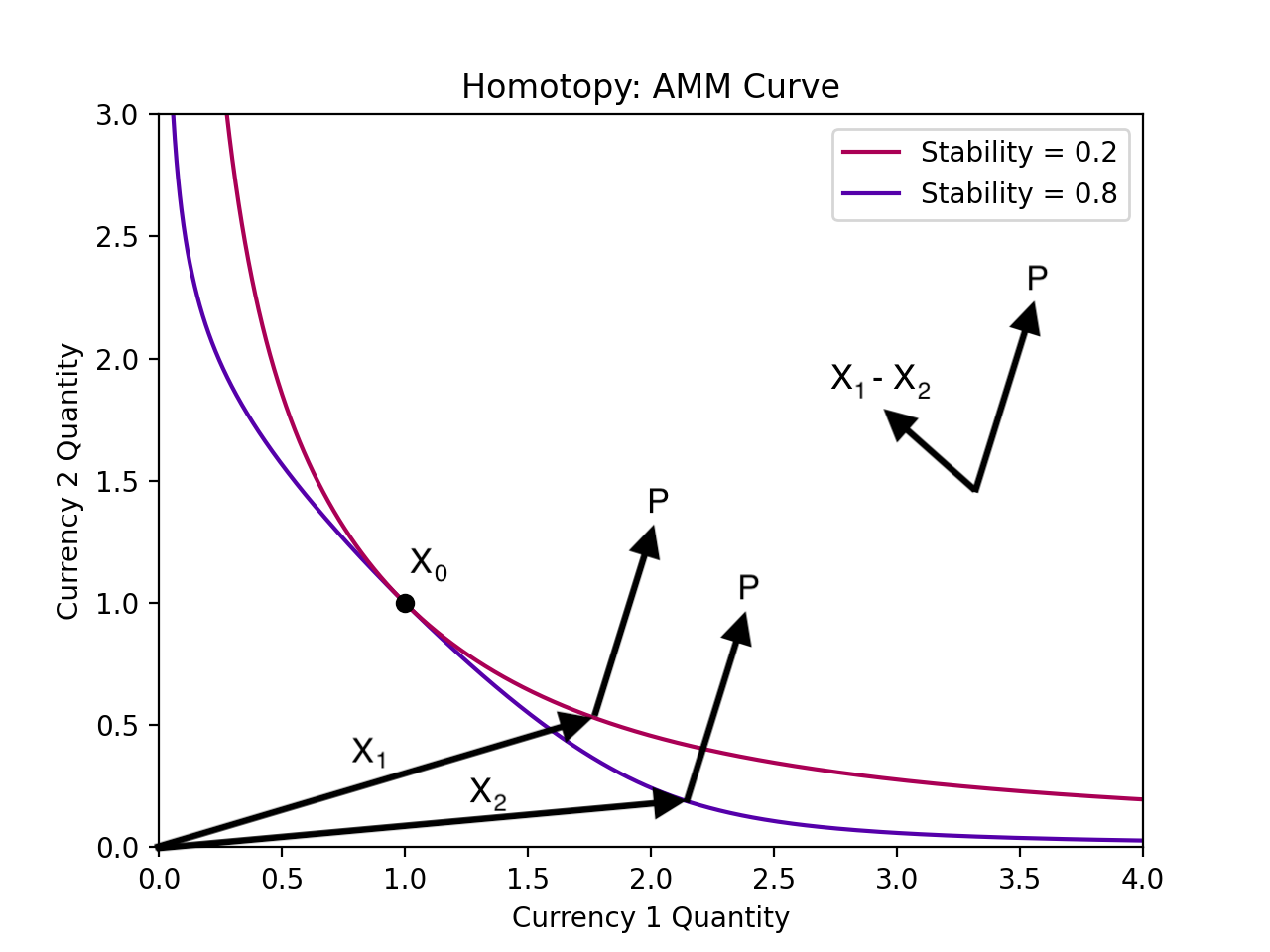}
\par\end{centering}
\caption{\label{fig: stability vs value}A visual demonstration of why lower
stability results in an increase in the portfolio value function.
$X_{1}$ and $X_{2}$ are the vectors pointing to the locations on
the AMM curves where the gradient is parallel to the price vector
$P$, and therefore these are the locations that minimize the remaining
value in the market. This figure shows that a decrease in stability
moves the minimizing point closer to the initial point $X_{0}$. The
claim is that the lower stability curve has a higher value, i.e. $P\cdot X_{1}>P\cdot X_{2}$.
This is equivalent to the condition that $P\cdot(X_{1}-X_{2})>0$,
and the pair of arrows in the upper right indicate that this is indeed
the case (since the angle between them is less than $90^{\circ}$).}

\end{figure}

As far as the mixing curves are concerned, the above Figure \ref{fig: portfolio value functions}
demonstrates the effects that stability has on the portfolio value
functions. The most notable phenomenon is the fact that the portfolio
value increases as the stability decreases. To see why this is the
case, note that Figure \ref{fig: stability vs value} shows the following
rather non-intuitive fact; given a fixed price vector $P$, the point
$X$ where $A_{t}^{hom}(X)=k$ and $\nabla A_{t}^{hom}(X)=\lambda P$
will be closer to the initial point $X_{0}$ when stability is lower.
In other words, one must be careful when thinking about the meaning
of stability. An increase in stability means that a change in quantity
vector $X$ results in a smaller change in the corresponding price
vector $P$. Conversely, a decrease in stability means that a change
in price vector $P$ results in a smaller change in the corresponding
quantity vector $X$. Because lower stability means higher slippage,
the takeaway here is that (1) an AMM with higher slippage will tend
to have higher portfolio value functions and (2) AMMs with greater
sensitivity to user behavior are better able to hold value. Further
work is needed to show that this notion holds true in a broader context,
but this and Figure \ref{fig: stability vs value} are helpful in
initial understanding.

\section{Conclusion}

This work presents a novel method for blending the benefits of constant
sum and constant product market makers. In particular, the non-uniform
homotopy construction of a CSMM-to-CPMM blend opens many doors for
designing interesting and intuitive AMM behavior. The homotopy construction
gives a very concrete geometric meaning to $t$ as the blending weight.
The $\lambda(s,t)$ homotopy parametrization allows for very exact
computations and prevents drift due to rounding errors. In addition,
thinking of $t$ as a function of $s$ allows for easy customizability
of non-uniform curves. More investigation is needed to see the full
behavior of these non-uniform curves, but already it is clear that
the balance between stability and slippage is significant factor in
designing market makers.

Future directions for this work include a variety of possible options.
There are many designs and behavior types to consider in designing
new non-uniform homotopy markets, especially in a higher dimensional
setting. There is plenty of room for more sophistication in the models
weighing the balance of stability and slippage. Determining how to
algorithmically and dynamically adjust the parameters of the AMM could
be helpful in maintaining stable behavior even under large external
price changes. Portfolio value function behavior as it relates to
AMM parameters like stability and bias should be investigated in more
depth. In short, this work lays a foundation for the initial design
of a new class of market makers with a wide variety of possible desired
behaviors.

\newpage{}


\begin{thebibliography}{AEC21}
\bibitem[AEC21]{AEC21}Angeris, Guillermo; Evans, Alex; Chitra, Tarun:
Replicating Market Makers. March 26, 2021 (accessed November 17, 2021)

\bibitem[Ego19]{Ego19}Egorov, Michael: StableSwap - efficient mechanism
for Stablecoin liquidity. November 10, 2019 (accessed November 5,
2021)

\bibitem[Ego21]{Ego21}Egorov, Michael: Automatic market-making with
dynamic peg. June 9, 2021 (accessed November 10, 2021)

\bibitem[EH21]{EH21}Engel, Daniel; Herlihy, Maurice: Composing Networks
of Automated Market Makers. August 31, 2021 (accessed November 18,
2021)

\bibitem[TPL22]{TPL22}Tiruviluamala, Neelesh; Port, Alexander; Lewis,
Erik: A General Framework for Impermanent Loss in Automated Market Makers. March 23, 2022 (arXiv:2203.11352)
\end{thebibliography}
\end{document}